\documentclass[10pt]{article}
\usepackage{fullpage}
\usepackage{setspace}
\usepackage{parskip}
\usepackage{titlesec}
\usepackage[dvipsnames]{xcolor}
\usepackage{verbatim}
 \usepackage{breakcites}
 \usepackage{lineno}
 \usepackage{hyphenat}
 \usepackage{placeins}

 \usepackage{times}

 \PassOptionsToPackage{hyphens}{url}
 \usepackage[colorlinks = true,
             urlcolor  = blue,
             citecolor = blue,
             anchorcolor = blue]{hyperref}
 \usepackage[style=chem-acs, articletitle=true, maxbibnames=10,backend=bibtex]{biblatex}
 \bibliography{kondo}

 \usepackage{xr-hyper}
 \usepackage{etoolbox}

 \titlespacing{\section}{0pt}{*3}{*1}
 \titlespacing{\subsection}{0pt}{*2}{*0.5}
 \titlespacing{\subsubsection}{0pt}{*1.5}{0pt}

 \usepackage{authblk}
\usepackage{sectsty}

\usepackage[ngerman,english]{babel}
\usepackage{graphicx}
\usepackage{amsfonts,amsmath,amssymb}

\usepackage{float}

\usepackage[version=4]{mhchem}

\usepackage{lineno}

\makeatletter
\newcommand*{\addFileDependency}[1]{
  \typeout{(#1)}
  \@addtofilelist{#1}
  \IfFileExists{#1}{}{\typeout{No file #1.}}
}
\makeatother

\usepackage{amsmath}
\usepackage{bm}
\usepackage{graphicx,color}
\usepackage{subfigure}
\usepackage{array}
\newcommand{\REV}[1]{\textcolor{.}{#1}}

\newcommand*{\veck}{{\mathbf{k}}}

\makeatletter
\def\@hangfrom@section#1#2#3{\@hangfrom{#1#2}#3}
\def\@hangfroms@section#1#2{#1#2}
\makeatother

\definecolor{myblue}{rgb}{0,0,1}

\begin{document}
\title{Towards an exact electronic quantum many-body treatment of Kondo correlation in magnetic impurities}
\author[1,2,3]{Tianyu Zhu \thanks{tianyu.zhu@yale.edu}}

\author[1,3]{Linqing Peng}

\author[1]{Huanchen Zhai}

\author[1]{Zhi-Hao Cui}

\author[1]{Runze Chi}

\author[1]{Garnet Kin-Lic Chan \thanks{gkc1000@gmail.com}}

\affil[1]{Division of Chemistry and Chemical Engineering, California Institute of Technology, Pasadena, CA, USA 91125}
\affil[2]{Department of Chemistry, Yale University, New Haven, CT, USA 06520}
\affil[3]{These authors contributed equally to this work}

\date{}

\maketitle

\doublespacing

\section*{Abstract}
The Kondo effect is a prototypical quantum  phenomenon arising from the interaction between localized electrons in a magnetic impurity and itinerant electrons in a metallic host. Although it has served as the testing ground for quantum many-body methods for decades, the precise description of Kondo physics with material specificity remains challenging. Here, we present a systematic \textit{ab initio} approach to converge towards an exact zero-temperature electronic treatment of Kondo correlations. Across a series of $3d$ transition metals, we extract Kondo temperatures matching the subtle experimental trends, with an accuracy exceeding that of standard models. We further obtain microscopic insight into the origin of these trends. More broadly, we demonstrate the possibility to start from fully \textit{ab initio} many-body simulations and push towards the realm of converged predictions. 

\newpage 

\section*{Introduction}
The Kondo system of a magnetic impurity embedded in a non-magnetic metallic host is a foundational  quantum many-body materials problem~\cite{Hewson1993a, Kondo1964,Knorr2002,Nevidomskyy2009,Pruser2011,Kouwenhoven2001, lucignano2009kondo}. As the temperature is decreased below a characteristic Kondo temperature $T_\mathrm{K}$, the impurity moment is screened by the conduction electrons, forming a many-electron singlet state that manifests as a sharp resonance in the local density of states. 
Although the physics involves many-electron correlations in a bulk material, a qualitative understanding was achieved decades ago 
through the Anderson impurity model (AIM)~\cite{Anderson1961} and via Wilson's numerical renormalization group (NRG) solution of the Kondo spin problem~\cite{Wilson1975b}.  However, while deeply insightful, these model frameworks are limited in their quantitative precision due to the uncertainties in the specific model parameters and the neglect of higher-energy electronic degrees of freedom. 
In this work, we show that one can now describe Kondo physics without ever simplifying to such models, starting only with the \textit{ab initio} many-electron Hamiltonian of the full material. In so doing, we present a precise material-specific treatment of the phenomenon, which can in principle be converged towards a numerically exact solution.

The challenge of the Kondo problem stems from the simultaneous presence of strong electron correlation leading to local singlet formation, and impurity-metal hybridization, which requires considering the thermodynamic limit. In addition, the Kondo energy scale is very small (typically 1 to 500 K).
In low-energy models, 
one chooses a (small) set of impurity correlated orbitals (e.g., a few $d$ or $f$ orbitals), whose interactions are formally described by a downfolded effective Hamiltonian. Uncertainties, which are large on the Kondo scale, then arise from
the choice and construction of these orbitals~\cite{Karp2021a},
the derivation of the effective interactions (and approximations to treat their complicated frequency dependence~\cite{Aryasetiawan2004b}),  
and the use of double-counting corrections to remove density functional theory (DFT) contributions~\cite{Wang2012,Muechler2022b}, in lieu of a fully many-body treatment of the local interactions. As a result, although much qualitative progress has been made in describing Kondo physics in different realistic settings~\cite{Jacob2009,Surer2012a,Gardonio2013,Dang2016a,Valli2020}, the \REV{accurate} simulation of Kondo trends 
across different magnetic impurities, geometries, or even different calculations, without reference to experimental data, is challenging~\cite{Surer2012a,Dang2016a,Gardonio2013}. 

We will instead attempt to solve the \textit{ab initio} many-electron Schr\"odinger equation for the Kondo problem without first deriving a low-energy model. 
This offers the advantage that it is often easier to quantify (and thus converge) errors in the approximate solution of a problem, than the error associated with deriving a model. 
We build on our work on full cell embedding in the context of dynamical mean-field theories~\cite{Zhu2020,Cui2020,Zhu2021c,Cui2022,Cui23-cuprate-doping}
to construct a systematically improvable representation of the impurity in its metallic host. Solving the many-body problem in this representation, as we increase the number of orbitals in the parent basis, we eventually obtain the exact non-relativistic 
description of the pure electronic problem. In small molecules~\cite{larsson2022chromium}, as well as in simple materials (such as organic crystals~\cite{yang2014ab}), it has been established that for the quantities of interest, related strategies can reach an accuracy rivalling, or even exceeding, that of experimental measurements. We thus ask, how far can we go with a similar philosophy for a realistic correlated quantum materials problem?

A major technical challenge is to solve for the impurity properties given the large number of electronic orbitals. We achieve this by implementing a highly-parallel quantum chemistry density matrix renormalization group (DMRG) \REV{eigenstate} and Green's function solver~\cite{Zhai2021a,Zhai2023a}. The use of delocalized (i.e., molecular orbital) representations reduces the entanglement and allows us to converge the many-body solvers. 
Within our framework, we simulate the series of seven 3$d$ magnetic impurities (Ti, V, Cr, Mn, Fe, Co, Ni) in a bulk copper host at zero \REV{and finite} temperatures, computing quantities such as the local density of states and quasiparticle (QP) renormalization, \REV{excited states}, orbital occupancies, and spin correlations. Using the QP renormalization and orbital occupancies as \REV{sensitive} metrics, we demonstrate convergence of our simulations to the parent basis solution, and estimate the remaining error in the parent basis. 
\REV{Extracting the Kondo temperature from the quasiparticle renormalization,}
the converged simulations capture subtle trends 
across the 3$d$ series as seen in experiments (Fig.~\ref{fig:dos}b), \REV{give new insights into the element-specific mechanisms,}
and improve predictions from models that use standard parametrizations by an order of magnitude or more.
They thus demonstrate the potential of approaches based on fully \textit{ab initio} simulations in the interpretation of complex correlated electron phenomena.



\section*{Results}

\begin{figure}[hbt!]
\centering
\includegraphics[width=0.9\textwidth]{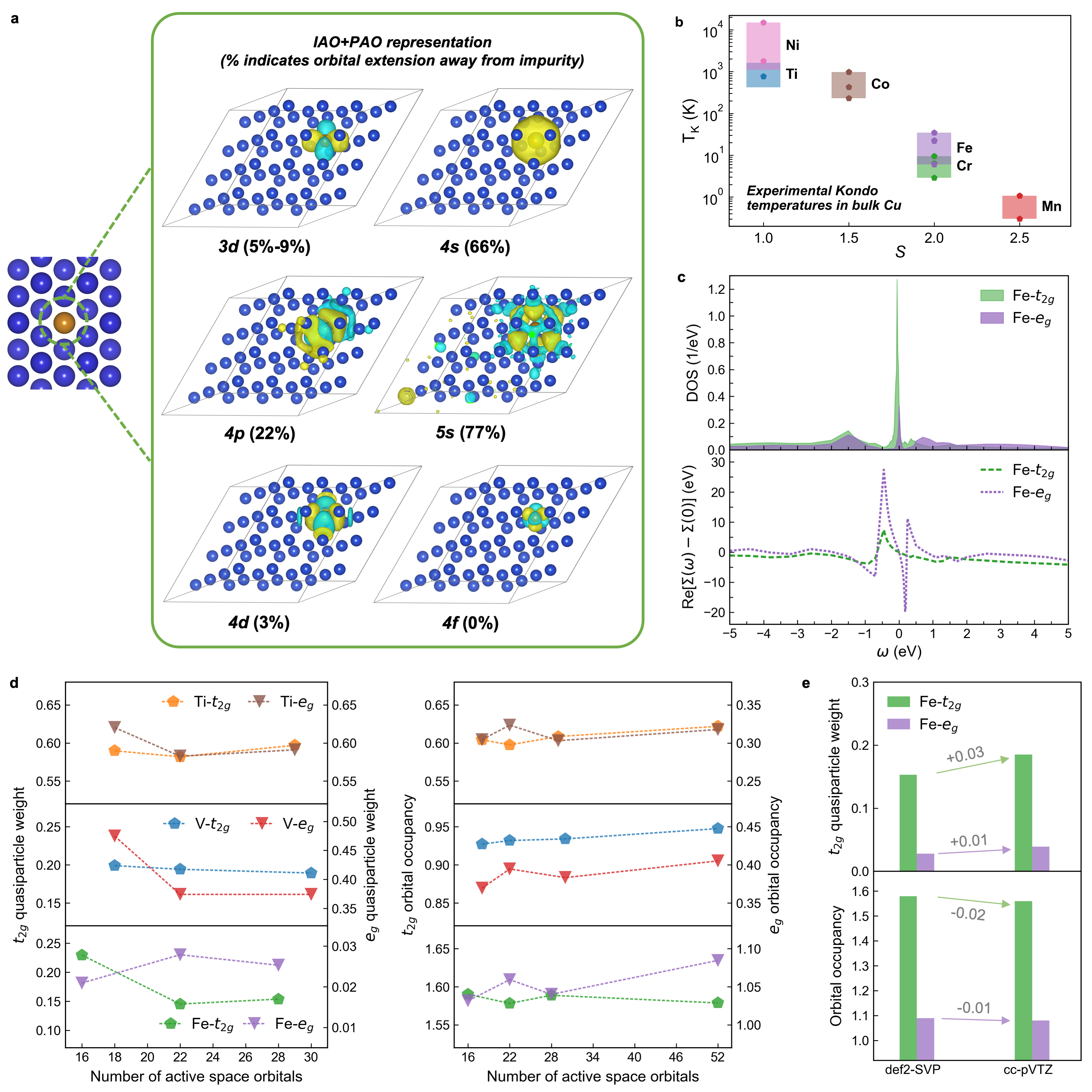}
\caption{Illustration and validation of computational strategy. (a) All-orbital quantum embedding strategy, where general Coulomb interactions between all valence and high-virtual orbitals of the impurity atom are treated at the quantum many-body level. The percentage of orbital extension onto neighbouring Cu atoms is estimated through a L\"owdin population analysis. (b) \REV{Experimentally measured Kondo temperatures for $3d$ transition metal impurities in bulk Cu. Each dot corresponds to a distinct measurement, while the shaded areas illustrate the range of Kondo temperatures from different observables and their associated measurement uncertainties (where applicable)~\cite{Daybell1968}. The $x$-axis shows the nominal spin ($S$) of the isolated impurity atoms.} (c) Simulated Kondo resonance and self-energies of the $t_{2g}$ and $e_g$ orbitals of the Fe impurity with an active space of (22e, 22o) in the def2-SVP basis. (d) Estimated convergence of the quasiparticle renormalization weights and orbital occupancies of the \REV{$t_{2g}$ and $e_g$ orbitals, averaged within each manifold,} of Ti, V, and Fe impurities as the DMRG active space increases in the def2-SVP basis. (e) Comparison of quasiparticle renormalization weights and orbital occupancies of $3d$ orbitals of the Fe impurity calculated in the def2-SVP basis versus the cc-pVTZ basis.}
\label{fig:dos}
\end{figure}

\paragraph{Numerical strategy.} 
Our basic plan is to describe the magnetic impurity atom with as complete a basis  of electronic orbitals as practical in an impurity-bath embedding setup (Fig.~\ref{fig:dos}a). 
As the impurity orbital space is increased, it formally converges to the Hilbert space of the material and thus to an exact electronic description (i.e., phonon effects are excluded); for any finite impurity space, the bath approximates the neglected degrees of freedom in the material. While achieving convergence for bulk properties would require including orbitals that span the electronic space of all atoms of the material, here we are focused on observables on the impurity, where convergence of an impurity-centered basis expansion is more rapid. Converging the representation in this way means that we do not need to first downfold the Hamiltonian to derive a model.

More specifically, we study the series of seven $3d$ transition metal impurities (Ti, V, Cr, Mn, Fe, Co, Ni) in bulk Cu, with associated Kondo temperatures that span three orders of magnitude. We started from a Gaussian atomic orbital representation of all atoms: a split-valence double-$\zeta$ polarization basis (def2-SVP)~\cite{Weigend2005} for the impurity atoms and a slightly smaller  (def2-SV(P)) basis for Cu (which omits the $4f$ basis shell). The def2-SVP basis for the impurity atoms contains the $1s2s2p3s3p3d4s4p4d4f5s$ shells (denoted a $5s3p2d1f$ basis, from the shell-count) and thus goes significantly beyond the $3d$ shell considered in a model Hamiltonian treatment. To test the convergence of the results, for a subset of the calculations we also used a larger cc-pVTZ basis~\cite{Balabanov2005a} on the impurity atom, corresponding to a $7s6p4d2f1g$ basis. This  further improves the electronic treatment of the impurity.  
We will refer to these as the parent bases and below we demonstrate convergence towards the exact solutions within the parent bases. 
We estimate the remaining error of the parent basis by the difference between the def2-SVP and cc-pVTZ results. 

Starting from DFT-optimized XCu\textsubscript{63} structures (X = impurity), we performed periodic DFT calculations in the Gaussian atomic orbital bases with the PBE functional~\cite{Perdew96PBE,Sun2020b}. 
The Gaussian basis functions were then transformed into an orthogonal basis of intrinsic atomic orbitals plus projected atomic orbitals (IAO+PAO)~\cite{Knizia2013d}. 
The impurity IAOs and PAOs are visualized in Fig.~\ref{fig:dos}a. 
The higher shell orbitals in this picture extend away from the impurity atom onto the neighbours, and because of the close packing of the atoms, start to capture important electronic degrees of freedom  of the atoms neighbouring the impurity. For example, using the L\"owdin population to measure the spatial extent of the $4s$ orbital of an Fe impurity, we find that it is close to $66\%$ on the neighbouring Cu atoms in FeCu\textsubscript{63}. The large basis sets may thus be viewed as forming an ``impurity-centered'' basis expansion, similar to well-studied local correlation treatments in quantum chemistry.

The impurity-bath embedding Hamiltonian takes the form
\begin{align}
    H_\mathrm{emb} = \sum_{ij}^\mathrm{imp} \tilde{F}_{ij} a_i^\dagger a_j + \frac{1}{2} \sum_{ijkl}^\mathrm{imp} (ij|kl) a_i^\dagger a_k^\dagger a_l a_j 
    + \sum_{ip} V_{ip} (a_i^\dagger c_p + c_p^\dagger a_i) + \sum_p \epsilon_p c_p^\dagger c_p 
    \label{eq:AIM}
\end{align}
where the impurity sum extends over all IAOs and PAOs in the impurity basis, $\{a^{(\dagger)}_i\}$ are creation/annihilation operators for impurity states, and $\{c^{(\dagger)}_{p}\}$ are creation/annihilation operators for bath states.
The impurity Coulomb interaction matrix $(ij|kl)$ is taken as the bare Coulomb interaction in the 
IAO+PAO basis, while the impurity one-particle interaction $\tilde{F}_{ij}$ in Eq.~\ref{eq:AIM} is defined as the Hartree-Fock effective Hamiltonian with the local mean-field potential subtracted (this subtraction is diagrammatically exact, so there is no double counting). 
\REV{All interactions are treated non-relativistically (spin-orbit coupling is beyond the scope of this work). }The impurity-metal hybridization function $\Delta(\omega)$ was obtained at the PBE level. Since we used a Hamiltonian-based impurity solver, we discretized the $3d4s$ valence block of the hybridization function on a non-uniform grid along the real frequency axis~\cite{DeVega2015c}
\begin{equation}
    \Delta_{ij}(\omega) = \sum_{p} \frac{V_{ip} V_{jp}}{\omega - \epsilon_p }, ~ij \in 3d4s
\end{equation}
where $\{\epsilon_p\}$ and $\{V_{ip}\}$ are the bath energies and impurity-bath couplings in Eq.~\ref{eq:AIM}. We used 49 bath orbitals to couple to each valence impurity orbital.
The total embedding problem thus consisted of 300 electrons in 316 (impurity plus bath) orbitals using the impurity def2-SVP basis, denoted (300e, 316o), and 300 electrons in 353 orbitals using the impurity cc-pVTZ basis, denoted (300e, 353o).

To compute and converge the impurity Green's function, we constructed a series of smaller orbital spaces (active spaces) that are subspaces of the full orbital space of the embedding problem. We then monitored the convergence of the \REV{eigenstate} and dynamical quantities as a function of the active space size.
The active spaces were defined to span natural orbitals (eigenvectors of the one-particle density matrix) of an (approximate) ground state of the impurity problem, taking  natural orbitals with the largest orbital entropy (a procedure which  has been shown to yield highly compact orbital subspaces in model impurity problems~\cite{Zgid2012}). 
We used active spaces of up to (36e, 52o) for the ground-state problem in the impurity def2-SVP basis and (46e, 76o) in the impurity cc-pVTZ basis, and up to (26e, 36o) for the dynamical properties and the finite temperature simulations.
We computed ground states and associated properties (e.g., density matrices and correlation functions), \REV{and as many as several hundred excited states,}  
using \textit{ab initio} DMRG~\cite{Chan2011a,Zhai2021a,Zhai2023a}. 
For dynamical properties, we 
used a new zero-temperature \textit{ab initio} dynamical DMRG (DDMRG) solver to compute the impurity self-energy on the real frequency axis~\cite{Ronca2017b,Zhai2023a}. (Benchmarks of the DDMRG solver on the single-impurity Anderson model are shown in Supplementary Note 2).

\paragraph{Convergence towards parent basis and an exact simulation.} 
We first consider the Fe impurity to illustrate general features of the results.
In Fig.~\ref{fig:dos}c we show the orbital-resolved excitation spectra and self-energies of the Fe $3d$ orbitals. As expected, we see  sharp Kondo resonance peaks, also known as  Abrikosov-Suhl resonances~\cite{tsvelick1983exact}, around the Fermi level for both the $t_{2g}$ and $e_g$ orbitals, with broad Hubbard shoulder peaks at higher energies. We find that the Fe-$e_g$ Kondo peak is narrower and smaller than the Fe-$t_{2g}$ peak from the stronger QP renormalization, which can also be seen from the real part of the real-axis self-energies shown in Fig.~\ref{fig:dos}c. The imaginary part of the real-axis self-energies of both the $t_{2g}$ and $e_g$ orbitals approaches zero at the Fermi level (Fig.~S5), i.e., $\mathrm{Im}\Sigma(T=0, \omega) \rightarrow 0$ as $\omega \rightarrow 0$. The splitting of the Hubbard peaks is roughly $2.5 \sim 3$~eV (in the range of the screened interaction for Fe $3d$ orbitals~\cite{Nakamura2008a}) and is slightly smaller for the $t_{2g}$ orbitals than for the $e_g$ orbitals.

We then examine the convergence of impurity observables towards the parent basis limit. In Fig.~\ref{fig:dos}d, we show the orbital-resolved QP renormalization weight $Z$ (calculated on the real axis as $
    \Big[1 - \frac{\partial \Sigma(\omega)}{\partial \omega} \Big|_{\omega=0} \Big]^{-1}$)
and the orbital occupancy $n$ of the $3d$ orbitals, for the Ti, V, and Fe impurities, as a function of active space size with a parent def2-SVP basis. Additional convergence data are in the Supplementary Notes 3 and 4, including the convergence with respect to the DMRG bond dimension (for which the uncertainty is substantially smaller than the uncertainty from the active space for all cases other than Mn).
We observe that the QP weight is clearly more challenging to converge. Examining the full series of elements (Supplementary Notes 3 and 4) we conservatively estimate that, with the exception of Mn,  $n$ is converged to at least $\sim 0.04$ and $Z$ to at least $\sim 0.05$ of the exact parent basis result. The latter exception arises because it was difficult to fully converge the DMRG calculations in the largest active spaces for Mn. 

Beyond these errors, the primary source of deviation from an exact simulation comes from the insufficiency of the parent basis itself. To estimate the error from the parent basis, in Fig.~\ref{fig:dos}e we show the change in $Z$ and $n$ as we increase the parent basis from def2-SVP to cc-pVTZ, for the Fe impurity. We see that both quantities change by less than $\sim 0.03$. Standard quantum chemistry arguments suggest that most quantities converge like the inverse cube of the cardinal number of the basis~\cite{helgaker1997basis} (2 in the case of def2-SVP, 3 for cc-pVTZ), which indicates that $Z$ and $n$ are well converged with respect to the representation of the electronic degrees of freedom of the Fe impurity and its close neighbours. Only the contribution from fluctuations involving long-range Coulomb interactions beyond the closest neighbours remains outside of our treatment and this error analysis.


\paragraph{\REV{Detailed Kondo temperature trends and mechanisms.}}

Having examined the convergence of our numerical results, we now study the trends in Kondo physics across the series of 7 elements. 
For $3d$ magnetic impurities in bulk Cu, experiments observe an exponential decay in Kondo temperatures moving from nominally low-spin (e.g., Ti) to high-spin (e.g., Mn) impurities~\cite{Daybell1968}. 
The estimated experimental Kondo temperatures for the $3d$ magnetic impurities, along with the nominal spins $S$ of the atom, are shown in Fig.~\ref{fig:dos}b. 
Note that the experimental estimates of $T_\mathrm{K}$ involve significant interpretation (for a full discussion see Ref.~\cite{Daybell1968}) and thus a range of characteristic temperatures are obtained depending on the type of measurement and method to extract $T_\mathrm{K}$. 

\REV{We estimate theoretical Kondo temperatures} 
from Hewson's renormalized perturbation theory of the Anderson impurity model~\cite{Hewson1993}
\begin{equation}
    T_\mathrm{K} = -\frac{\pi}{4} Z \cdot \mathrm{Im} \Delta(0).
    \label{eq:hewson}
\end{equation}
Here $\Delta(0)$ is the hybridization function at the Fermi level and $Z$ is the QP renormalization weight on the real axis (Fig.~S4 and Table~S6). \REV{Hewson's formula gives separate estimates of $T_\mathrm{K}$ for the $e_g$ and $t_{2g}$ orbitals}; the lower of the two Kondo temperatures is the relevant one for the comparison to measurements (see, e.g., Ref.~\cite{daybell1967observation} for resistivity). 

\begin{figure}[hbt!]
\centering
\includegraphics[width=\textwidth]{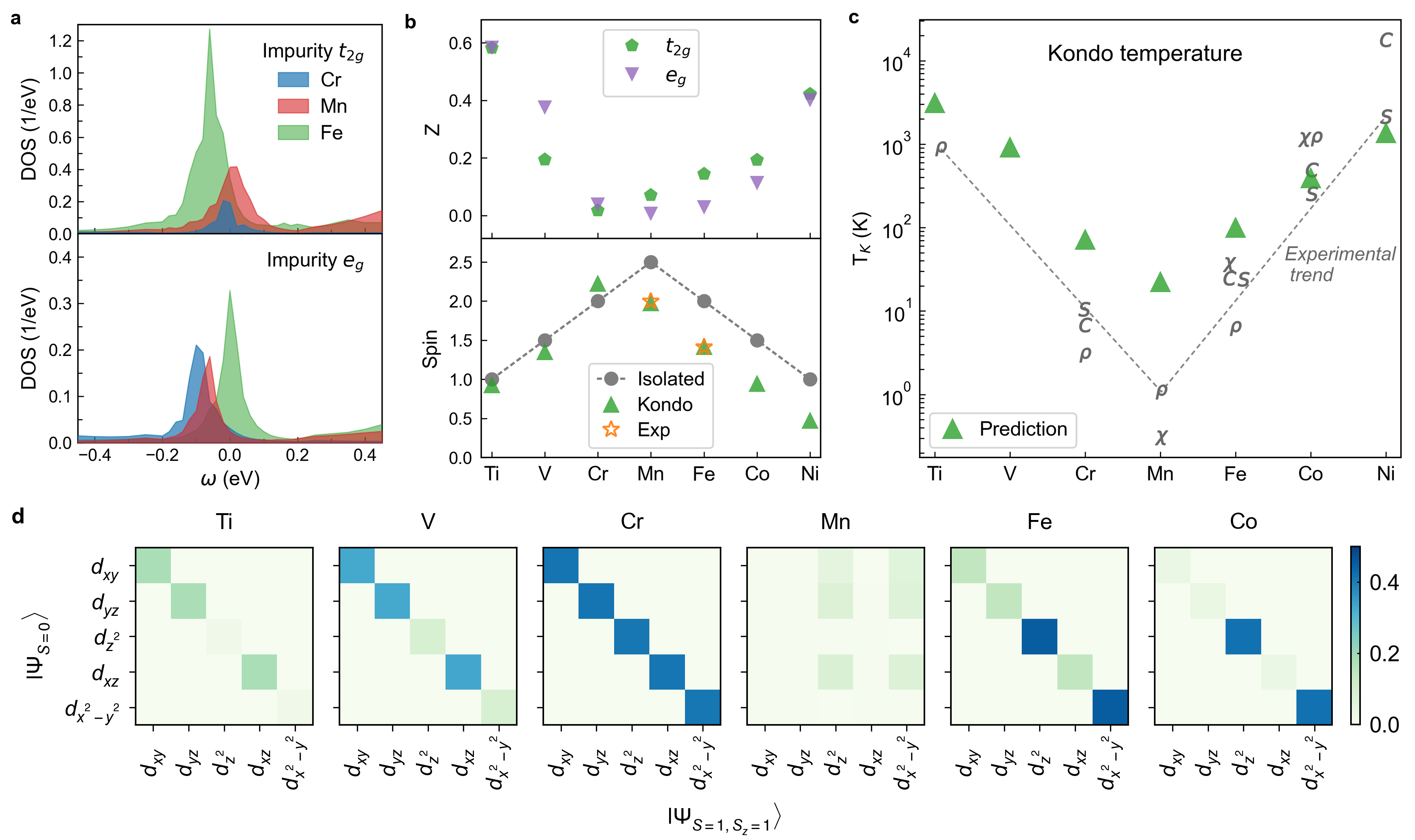}
\caption{Trends in impurity DOS, quasiparticle renormalization, spin moment, Kondo temperature, \REV{and orbital character of the lowest triplet excited state}. All results are in the def2-SVP basis with ground-state active space of (36e, 52o) and Green's function active space of (22e, 22o). (a) Comparison of DOS of $3d$ orbitals of Cr, Mn, and Fe impurities. (b) Top: Orbital-resolved quasiparticle renormalization weights $Z$. Bottom: Local spin moments of impurity atoms compared to nominal spins of isolated atoms and experimental values. (c) Kondo temperatures from all-orbital simulations, estimated by zero-temperature Green's function calculations using Hewson's formula, compared to experimental values~\cite{Daybell1968}. The experimental trend lines are provided as a visual aid. Symbols for each type of measurement are: $\rho$: resistivity, $\chi$: susceptibility, $C$: specific heat, $S$: thermoelectric power. \REV{(d) Orbital-resolved spin-flip transition density matrix for the lowest triplet excited state $|\langle \Psi_{S=0} | c^\dagger_{i\downarrow} c_{j\uparrow} | \Psi_{S=1, S_z=1}\rangle|$ within the impurity $3d$ subspace.}}
\label{fig:Tk}
\end{figure}

Our predicted Kondo temperatures for all the studied elements are shown in Fig.~\ref{fig:Tk}c. 
\REV{We capture the trend in $T_\mathrm{K}$ across the full set of elements}
in good agreement with the experimental estimates, \REV{and reproduce subtle differences such as the relative ordering of the lowest Kondo temperatures $T_\mathrm{K}(\mathrm{Cr})$, $T_\mathrm{K}(\mathrm{Mn})$, $T_\mathrm{K}(\mathrm{Fe})$.}
With the exception of Cr and Mn, \REV{the range of} our predicted $T_\mathrm{K}$ is within a factor of 2$\sim$3 of at least one of the experimental estimates of $T_\mathrm{K}$. As our calculations do not include long-range Coulomb fluctuations (beyond those captured by the basis extending to neighbouring atoms), this suggests that they do not contribute significantly at this level of accuracy. 
On the other hand, Mn, which has the lowest $T_\mathrm{K}$, is a system where we could not achieve numerical convergence of our results to the parent basis. This is the likely origin of the larger quantitative discrepancy \REV{for Mn}.

\REV{Conventionally, the origin of the observed trend in $T_\mathrm{K}$ is ascribed to  strong Hund's coupling in the impurity spins. In particular, when we represent the system by a multi-spin Kondo model, in the strong coupling limit the impurity spins are locked, which reduces the space of fluctuations and the effective exchange coupling $J_\mathrm{eff}$~\cite{okada1973singlet,Georges2013} (see Supplementary Note 15 for a brief summary).} Fixing the metallic host, and assuming all other effects are the same for different impurities, this Hund's driven relation can be simplified to
\begin{equation}
    T_\mathrm{K} \propto  \exp(-1/J_\mathrm{eff}) \propto \exp(-S),
    \label{eq:Tkspin}
\end{equation}
where $S$ is the spin of the magnetic atom. 
\REV{The above theoretical form gives qualitative agreement with the experimental data, as shown by the trend in Fig.~\ref{fig:dos}b. However, it also 
misses quantitative differences between certain magnetic impurities.} For example, it does not explain why the $T_\mathrm{K}$ of Cr (3$\sim$10 K) is lower than the $T_\mathrm{K}$ of Fe (30 K), if we consider both Cr and Fe atoms to have nominal $S=2$. \REV{Given that our \textit{ab initio} simulations faithfully reproduce the experimental trend in Kondo temperatures, we can now examine the Hund's suppression mechanism within our calculations.}

We start with the 
computed local ground-state properties of the impurity, such as the $t_{2g}$ and $e_g$ orbital occupancies across the impurity elements (Table~S7) and the effective spin (extracted from $\langle S^2\rangle$) in Fig.~\ref{fig:Tk}b. \REV{(Fig.~\ref{fig:Tk}a further illustrates the $t_{2g}$ and $e_g$ resolved DOS for Cr, Mn, Fe).}
Charge transfer takes place to
the impurity atom in all systems, giving negatively charged $3d$ shells compared to the isolated impurities. For instance, we find Fe in bulk Cu to be on average in a $d^7$ occupancy ($n_{\mathrm{Fe}}(3d)=6.92$), in excellent agreement with recent experimental estimates~\cite{Joly2017}. The strong QP renormalization in Fig.~\ref{fig:Tk}b tracks the partially-filled nature of the impurity $3d$ orbitals. For example, we obtain $n(t_{2g})=0.99$ in Cr and $n(e_g)=1.03$ in Mn (i.e., very close to half filling), which correlates well with their very small QP weights.

As a result of the charge transfer, the local spin moments of the magnetic impurities differ from their isolated-atom values (Fig.~\ref{fig:Tk}b). For Mn, Fe, Co, and Ni, we predict $S=1.99, 1.42, 0.95, 0.48$, about 0.5 lower than the isolated-atom spins. These $S$ values agree well with the available experimental data for Mn and Fe~\cite{Daybell1968,Haen1976,Joly2017}. Cr has the highest spin moment ($S=2.23$) among all $3d$ transition metal impurities, which deviates from the common understanding of this Kondo series~\cite{Nevidomskyy2009}, but is also supported by experimental measurements that suggest Cr is a high-spin impurity ($S=5/2$)~\cite{Haen1976,Sacramento1990}. 

\REV{Taking into account the quantitative differences in the observed $S$ in the Kondo systems and the isolated atoms, our results partially confirm a Hund's driven suppression mechanism. With the exception of the relative ordering of $T_\mathrm{K}(\mathrm{Cr})$ and $T_\mathrm{K}(\mathrm{Mn})$, higher observed $S$ indeed correlates with lower $T_\mathrm{K}$. 
For example, accounting for the charge transfer to Fe and Cr that modifies their effective spin, we can predict the observed relation $T_\mathrm{K}(\mathrm{Cr}) < T_\mathrm{K}(\mathrm{Fe})$. 
The Hund's picture can further be confirmed by examining the nature of the lowest triplet excited states which are associated with the destruction of the Kondo singlet. As seen in Fig.~\ref{fig:Tk}d, for most of the elements, the main character of the excitation is a coherent spin-flip of all the (close to) singly occupied $d$ orbitals, as expected in the limit of strong Hund's coupling.} 

\REV{The exception to this is Mn. Note that from the computed effective spin, the Hund's mechanism would predict Cr to have a {lower} $T_\mathrm{K}$ than Mn, contrary to the experimental findings. However, after charge transfer, the Mn ion has an excess (0.66) electron in the $t_{2g}$ shell, which experiences an unfavourable repulsion. 
We find that the lowest triplet excitation  corresponds to a partial transfer of an electron from $t_{2g}$ to the bath, accompanied by a spin-flip (see Supplementary Note 13). The low energy of this triplet state recalls the double-exchange mechanism~\cite{anderson1955considerations}, which favours same-spin alignment to enable charge delocalization. In fact, in our calculations, other than for Cr (which has a close to exactly half-filled $d$-shell when embedded in Cu) this type of excitation occurs at low energy in all the elements, but is especially low in energy for Mn. The inclusion of orbital degrees of freedom is thus an important part of the systematic trends in the Kondo temperatures, not only through the effective spin, but also through fluctuations.}


\REV{Some additional insight is provided by examining the detailed contributions to Hewson's formula. Because it starts from a Fermi liquid it naturally captures some effects omitted in the Kondo model, such as the effect of charge fluctuations and the crystal field splitting. Generally, the lower of the $T_\mathrm{K}(t_{2g})$ and $T_\mathrm{K}(e_g)$  estimates corresponds to the shell with the most unpaired electron character, consistent with a Kondo model where the impurity is replaced by a spin. However, for impurities where both $t_{2g}$ and $e_g$ have significant unpaired electron character (such as in Cr) the Hewson formula further identifies which shell gives rise to the lower Kondo temperature.} We find that the Kondo energy scales of Mn, Fe, Co, and Ni are \REV{associated with} $e_{g}$, while they are \REV{associated with} $t_{2g}$ for Cr, V, and Ti. As $\Delta(0)$ only varies slightly across the series (from $-0.59$ eV (Ti) to $-0.43$ eV (Fe) for the $t_{2g}$ orbital and from $-0.64$ eV (Ti) to $-0.38$ eV (Ni) for the $e_g$ orbital), \REV{the trend is generated primarily by the QP renormalization weight, which contains both the  Hund's coupling mechanism as well as the effect of charge fluctuations.}

\begin{figure}[hbt!]
\centering
\includegraphics{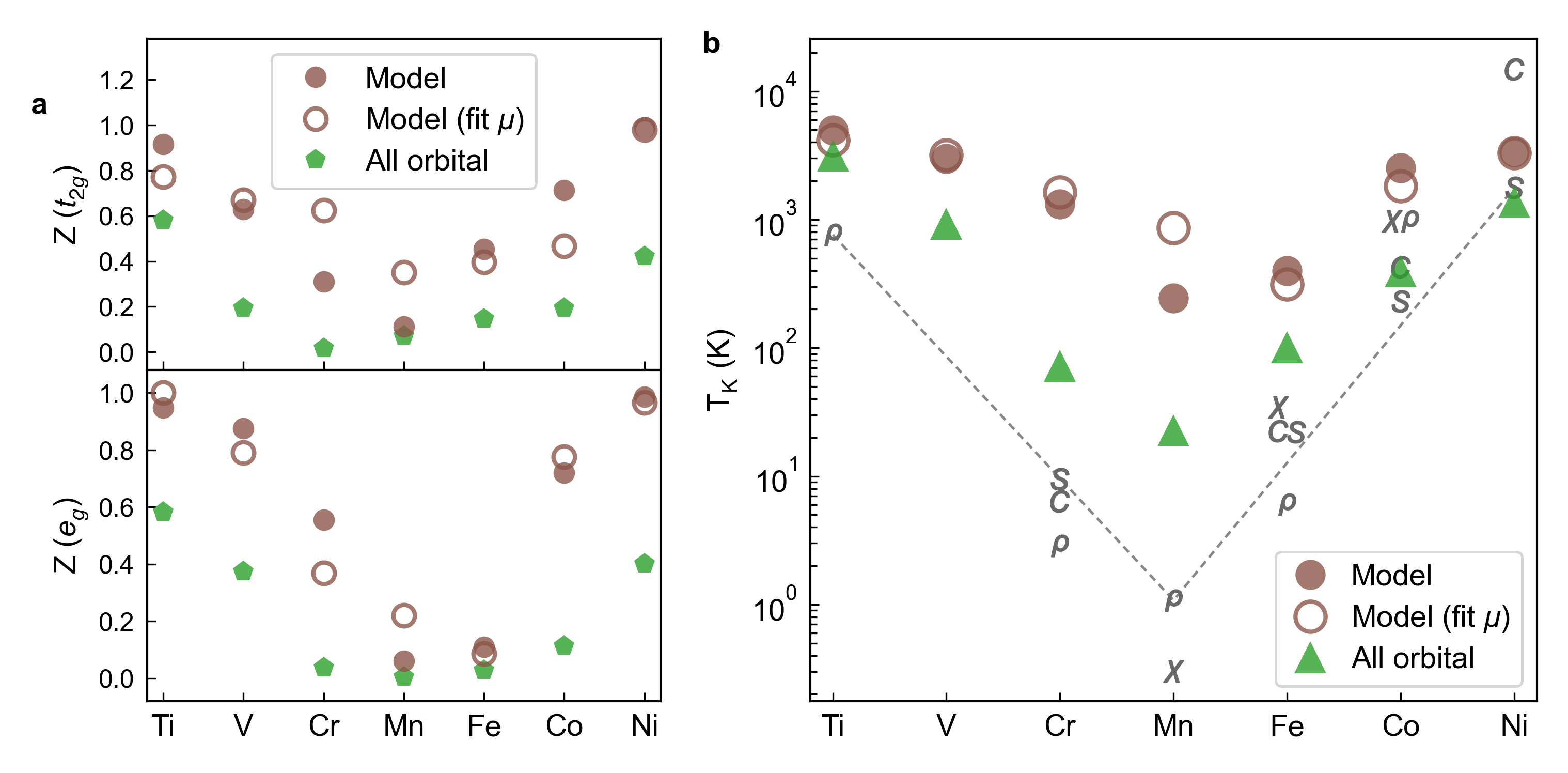}
\caption{Comparison \REV{ of \textit{ab initio} all-orbital simulations with model simulations using either the same chemical potentials (``Model'') or fitted chemical potentials to match the total $3d$ orbital occupancy of the all-orbital simulations (``Model (fit $\mu$)'')}. (a) Orbital-resolved quasiparticle renormalization
weights. (b) Predicted Kondo temperatures compared to the experimental estimates.}
\label{fig:model}
\end{figure}

\paragraph{Comparison to low-energy models.} Finally, to place the fidelity of our results in perspective, we check to see if the above predictions are easily captured within a standard application of the downfolded model approach. For this, we employ a five-orbital Anderson impurity model using 
the Kanamori Hamiltonian~\cite{Kanamori1963a,Georges2013} with screened Coulomb interaction and exchange parameters $U$ and $J$ taken from constrained random phase approximation (cRPA) calculations~\cite{Sasioglu2011}. (Note these parameters are also very close to previous parameters used in other model Kondo studies, e.g., for Co in Cu~\cite{Surer2012a,Valli2020}). We employed the same valence hybridization, bath discretization, and active-space DMRG solvers described above. \REV{The model was tested using both the same chemical potentials as in the all-orbital calculations, and refitted ones to match the total $3d$ orbital occupancies in the all-orbital results. Further} details and analysis are given in Supplementary Notes \REV{10 and 
11}. Fig.~\ref{fig:model} shows the estimated Kondo temperatures and quasiparticle renormalizations obtained from this model. 
We find that the quasiparticle renormalization weights of most magnetic impurities are significantly overestimated compared to our all-orbital simulations. As a result, the predicted $T_\mathrm{K}$'s do not even show a clear exponential trend, with a large overestimation of $T_\mathrm{K}$. In addition, the relative $T_\mathrm{K}$'s of \REV{Cr, Mn, and Fe}, are not reproduced \REV{using either the same chemical potential as in the \textit{ab initio} all-orbital calculations, or the refitted chemical potential.}
The quantitative improvement in the Kondo temperatures moving from the \REV{model to \textit{ab initio}} calculations \REV{is as much as 2 orders of magnitude.}
\REV{Our standard model includes 
the density-density, spin-flip, and pair-hopping interactions in rotationally invariant form.
In principle, a model that allows for all possible terms in the $3d$ shell, together with their full frequency dependence, should yield better estimates of the Kondo temperatures~\cite{Valli2020}. However, determining such a large number of downfolded parameters is challenging.}
 

\paragraph{Discussion.} We have demonstrated the predictive power of \textit{ab initio} quantum many-body simulations in the prototypical Kondo physics materials problem, namely, that of $3d$ transition metal impurities in bulk copper. By converging the material description and many-body treatment, we could reach an accuracy for the Kondo temperature that captures the subtle experimental trends across the $3d$ transition metal series.

In contrast to previous approaches, we achieved this accuracy by performing the quantum many-body simulations of the bare electronic problem with all orbitals, rather than within a downfolded model. 
This allows for a straightforward control of many aspects of converging our results towards an exact description of the phenomena. 
Within the zero-temperature electronic picture of our work, the physics of very long-range Coulomb fluctuations remain as an uncontrolled uncertainty, although these effects appear to be small on the scale of our results. 
However, our computational framework does not lie in opposition to model approaches. In particular, \REV{we showed how, after establishing the faithfulness of the physics within our framework, we can extract understanding in terms of traditional low-energy pictures.}
\REV{Moving forwards, the accuracy achieved here suggests that with continuing advances, we may move beyond qualitative descriptions of correlated electron phenomena, towards a systematically improvable simulation of observables directly measured by experiments.}



\section{Data availability}
Data used in this work are available in the main text and supplementary materials. 

\section{Code availability}
The fcDMFT code is available at \href{https://github.com/ZhuGroup-Yale/fcdmft}{https://github.com/ZhuGroup-Yale/fcdmft}. The libDMET code is available at \\\href{https://github.com/gkclab/libdmet\_preview}{https://github.com/gkclab/libdmet\_preview}. The block2 code is available at \href{https://github.com/block-hczhai/block2-preview}{https://github.com/block-hczhai/block2-preview}. PySCF is available at \href{https://github.com/pyscf/pyscf}{https://github.com/pyscf/pyscf}.

\section{Acknowledgements}
In the initial phase of this project performed at Caltech, T.Z. was supported by the Air Force Office of Scientific Research through the MURI program, Grant No. FA9550-18-1-0095.
The development of integral infrastructure by Z.-H.C. was supported by the US Department of Energy, Office of Science, Basic Energy Sciences, through award DE-SC0018140.
L.P.'s continuation of the work by T.Z. was supported by the Center for Molecular Magnetic Quantum Materials, an Energy Frontier Research Center
funded by the U.S. Department of Energy, Office of Science, Basic Energy Sciences under Grant No. DE-SC0019330.
T.Z. acknowledges support from the Air Force Office of Scientific Research under award number FA9550-24-1-0096 for the work conducted at Yale. H.Z. (DMRG solver) was supported by the US National Science Foundation, through Grant No. CHE-2102505. 
GKC is a Simons Investigator in Physics. Calculations were conducted in the Resnick High Performance Computing Center, supported by the Resnick Sustainability Institute at Caltech, as well as in the Yale Center for Research Computing. 

\section{Author Contributions}
T.Z., L.P., and G.K.-L.C. designed the initial project and wrote the manuscript. T.Z., L.P., and H.Z. developed the all-orbital DMFT method and code. T.Z., L.P., and Z.-H.C. performed \textit{ab initio} calculations of Kondo systems. T.Z. carried out the model Hamiltonian calculations. H.Z. developed the code of the DMRG impurity solver. \REV{R.C. provided additional theoretical analysis.} G.K.-L.C. supervised the project. All authors contribute to the discussion of the results as well as the writing and editing of the manuscript.

\section{Competing interests}
G.K.-L.C. is a part owner of QSimulate.

\section{Additional information}
Methods, Figures S1-S13, Tables S1-S13. 


\printbibliography 

\raggedbottom

\clearpage
\begin{center}
\textbf{\Large Supplementary Materials for: \\ Towards an exact electronic quantum many-body treatment of Kondo correlation in magnetic impurities}
\end{center}
\setcounter{equation}{0}
\setcounter{figure}{0}
\setcounter{table}{0}
\setcounter{page}{1}
\makeatletter

\setcounter{secnumdepth}{1}
\renewcommand{\theequation}{S\arabic{equation}}
\renewcommand{\thefigure}{S\arabic{figure}}
\renewcommand{\thetable}{S\arabic{table}}
\renewcommand{\thesection}{Supplementary Note \arabic{section}}

\sectionfont{\fontsize{12}{15}\selectfont}
\begin{refsection}
\section{Computational methods}
We generated initial XCu\textsubscript{63} (X = Ti, V, Cr, Mn, Fe, Co, Ni) structures by replacing one Cu atom in a $4\times4\times4$ supercell of bulk Cu with an impurity atom. We carried out DFT geometry relaxations for all XCu\textsubscript{63} structures with the PBE functional and projector-augmented-wave (PAW) basis using the Vienna Ab initio Simulation Package (VASP)~\cite{Kresse1996,Kresse1999}. The calculations were performed with a plane wave cutoff of 400 eV and a $\Gamma$-centered $3\times 3 \times 3$ $\mathbf{k}$-mesh. The forces on each atom were converged to less than 0.01 eV/\AA . With DFT-optimized structures, we performed single-point DFT calculations with the PBE functional in a periodic Gaussian basis set using the PySCF quantum chemistry software package~\cite{Sun2020b}. We employed a split-valence double-$\zeta$ polarization basis (def2-SVP) for all impurity atoms and a split-valence double-$\zeta$ basis (def2-SV(P)) for Cu~\cite{Weigend2005}. The correlation-consistent polarized triple-$\zeta$ basis (cc-pVTZ) was also employed for the Fe impurity to test the parent basis convergence~\cite{Balabanov2005a}. The libDMET code~\cite{Cui2020,Cui2022} was used to transform mean-field Fock matrix, density matrix, and electron repulsion integrals to the intrinsic atomic orbital plus projected atomic orbital (IAO+PAO) basis~\cite{Knizia2013d}. $3d4s$ atomic orbitals of all metal atoms were used as the predefined valence (minimal) orbitals in the IAO+PAO construction. 

We incorporate the full impurity atom into the embedding problem. In the IAO+PAO basis, the $3d4s4p4d4f5s$ (def2-SVP) and $3d4s4p4d4f5s5p5d5f5g6s6p6d7s7p$ (cc-pVTZ) orbitals of the impurity atoms were treated in the many-body impurity solvers, while the $1s2s2p3s3p$ orbitals were frozen at the mean-field level and left out of the embedding problem. The fcDMFT code~\cite{Zhu2019,Zhu2020,Zhu2021c} was used to perform the all-orbital calculations. We employed the bare Coulomb interaction $(ij|kl)$ within the impurity as the two-particle interaction matrix in the embedding Hamiltonian in Eq.~\ref{eq:AIM}. The one-particle impurity interaction matrix in Eq.~\ref{eq:AIM} is defined as
\begin{equation}
    \tilde{F}_{ij} = F^\mathrm{imp}_{ij} - \sum_{kl \in \mathrm{imp}} \gamma^\mathrm{imp}_{kl} [(ij|lk) - \frac{1}{2} (ik|lj)],
\end{equation}
where $F^\mathrm{imp}$ is the impurity Fock matrix computed by Hartree-Fock using the PBE density, and $\gamma^\mathrm{imp}$ is the impurity block of PBE density matrix. This definition ensures that there is no double counting in the impurity Hamiltonian.

The hybridization function for each impurity atom was obtained at the PBE level using a $4\times4\times4$ $\veck$-point sampling of the XCu\textsubscript{63} supercell:
\begin{equation}
    \Delta(\omega + i\delta) = \omega + i\delta - F_\mathrm{imp} - G_\mathrm{imp}^{-1}(\omega + i\delta),
\end{equation}
where $\delta$ is the broadening factor taken to be $\delta=0.01$ Ha. Since Hamiltonian-based impurity solvers (that require a bath) were employed, we discretized the $3d4s$ block of the hybridization function on a non-uniform grid along the real frequency axis using a pole-merging direct discretization method~\cite{DeVega2015c} to obtain bath energies $\{\epsilon_p\}$ and impurity-bath couplings $\{V_{ip}\}$ in Eq.~\ref{eq:AIM}, and 49 bath orbitals were coupled to each of the $3d4s$ valence orbitals. Among these 49 bath orbitals, 5 orbitals were placed within a $\mu\pm 0.027$ eV energy window, 18 orbitals were placed within the $[\mu \pm 0.027, \mu \pm 0.6]$ eV energy window, and the remaining 26 orbitals were placed at a higher energy window of $[\mu \pm 0.6, \mu\pm 8.0]$ eV. In total, the embedding problem contained 22 (impurity) + 294 (bath) = 316 orbitals in the def2-SVP impurity basis calculations. The same bath discretization procedure was applied to the cc-pVTZ impurity basis calculation of the Fe impurity, resulting in an embedding problem of 59 (impurity) + 294 (bath) = 353 orbitals.

To solve the embedding Hamiltonian, we first performed a Hartree-Fock calculation with the chemical potential fixed at the supercell DFT value, to define the number of electrons in the embedding problem. We note that the value of the chemical potential is often chosen to tune the orbital fillings, but we found that the current strategy gave excellent agreement with experimental occupancies of Mn and Fe. Following the HF solution, we carried out a configuration interaction with singles and doubles (CISD) calculation and computed natural orbitals by diagonalizing the CISD one-particle density matrix. A (36e, 52o) natural-orbital active space was then derived in the def2-SVP impurity basis calculations ((36e, 46o) for Ni, (46e, 76o) in the cc-pVTZ impurity basis calculation of Fe), where all kept natural orbitals have eigenvalues $n_i$ that satisfy $\mathrm{min}(n_i, 2-n_i) > 10^{-7}$. An \textit{ab initio} quantum chemistry DMRG calculation was then conducted on this active space to obtain ground-state properties including one-particle and two-particle density matrices and spin-spin correlation functions. The DMRG calculation was done with a bond dimension of $M=3500$ ($M=4000$ for Mn and Fe) using the block2 code~\cite{Zhai2021a,Zhai2023a}, where the discarded weight was below $2\times10^{-5}$ in all ground-state DMRG calculations. We further derived a series of smaller natural-orbital active spaces from the DMRG density matrix (see Supplementary Note 3 for details). 
The dynamical DMRG (DDMRG)~\cite{Ronca2017b} calculation was carried out for the smaller active spaces with a bond dimension up to $M=1500$ along the real axis at zero temperature. Larger bond dimensions of up to $M=4000$ were employed for strongly correlated sites to ensure that the discarded weight in the DDMRG calculations was below 0.02. To accommodate the non-uniform bath discretization, we used a broadening factor of $\eta=0.02$ eV within the $\mu \pm 0.25$ eV energy window, $\eta=0.05$ eV within the $[\mu \pm 0.25, \mu\pm 0.5]$ eV energy window, $\eta=0.2$ eV within the $[\mu \pm 0.5, \mu\pm 2.0]$ eV energy window, and $\eta=0.6$ eV within the $[\mu \pm 2.0, \mu\pm 5.0]$ eV energy window in the DDMRG calculations.

From the DDMRG calculations, we extracted the self-energy of the active space $\Sigma_\mathrm{DMRG, act}(\omega)$ as
\begin{equation}
    \Sigma_\mathrm{DMRG, act}(\omega) = G_\mathrm{DFT,act}^{-1}(\omega) - G_\mathrm{DMRG,act}^{-1}(\omega),
\end{equation}
where $G_\mathrm{DFT,act}$ was calculated from the effective DFT Hamiltonian rotated to the active space. The active-space self-energy was then rotated back to the full embedding space
\begin{equation}
    \Sigma_\mathrm{DMRG, emb}(\omega) = C^\mathrm{emb,act} \Sigma_\mathrm{DMRG, act}(\omega) (C^\mathrm{emb,act})^\dagger ,
\end{equation}
where $C^\mathrm{emb,act} = C^\mathrm{CISD} C^\mathrm{DMRG}$. Here, $C^\mathrm{CISD}$ represents the rotation matrix from the full embedding space to the natural-orbital active space derived from CISD calculations, while $C^\mathrm{DMRG}$ represents the rotation matrix from the CISD active space to the natural-orbital active space derived from ground-state DMRG calculations. Finally, the local Green's function of the impurity was calculated from Dyson's equation
\begin{equation}
    G_\mathrm{loc}(\omega) = [G_\mathrm{DFT, emb}^{-1}(\omega) - \Sigma_\mathrm{DMRG, emb}(\omega)]^{-1}.
\end{equation}

\section{Benchmark of DDMRG solver on Anderson impurity model} \label{sec:SIAM}
We benchmark the accuracy of the active-space DMRG and dynamical DMRG (DDMRG) solvers on a single-impurity Anderson model (SIAM), the fundamental model of Kondo physics, where high-accuracy numerical results are available (e.g., from NRG~\cite{Bulla1998} or DMRG~\cite{Nishimoto2004a}). The Hamiltonian of the SIAM is
\begin{equation}
    H = \sum_\sigma \epsilon_f f^\dagger_\sigma f_\sigma + U f^\dagger_\uparrow f_\uparrow f^\dagger_\downarrow f_\downarrow + \sum_{k \sigma} \epsilon_k c^\dagger_{k\sigma} c_{k\sigma} + \sum_{k\sigma} V_k (f^\dagger_\sigma c_{k\sigma} + c^\dagger_{k\sigma} f_\sigma),
\end{equation}
where $f^{(\dagger)}_\sigma$ are creation/annihilation operators for impurity states with spin $\sigma$ and energy $\epsilon_f$, $c^{(\dagger)}_{k\sigma}$ are creation/annihilation operators for band states with spin $\sigma$ and energy $\epsilon_k$, $U$ is the impurity on-site Coulomb interaction, and $V_k$ are the $k$-dependent coupling between impurity and band states. We followed Ref.~\cite{Bulla1998} and employed the flat-band hybridization function: 
\begin{equation}
    \mathrm{Im} \Delta(\omega+i0^+)=-0.015 D, ~ |\omega| < D
\end{equation}
where $2D$ is the conduction electron bandwidth. For convenience, we set $D=1$. The flat-band hybridization on a logarithmic grid was then discretized along the real axis to obtain $\{\epsilon_k, V_k\}$:
\begin{equation}
    -\mathrm{Im} \Delta(\omega+i0^+) = \sum_k \frac{V_k^2}{\omega-\epsilon_k}.
\end{equation}
We coupled 39 bath orbitals to the single impurity orbital. Furthermore, only the half-filling case was considered, which means $\epsilon_f=-\frac{1}{2}U$. We tested our impurity solver on three interaction strengths: $U=0.1, 0.2, 0.3$. 

We first solved the SIAM Hamiltonian using the Hartree-Fock approximation \REV{to determine an initial set of orbitals}. Within this (untruncated basis) we subsequently carried out a ground-state configuration interaction with single and double excitations (CISD) calculation or a ground-state DMRG calculation. Using these, we then derived CISD \REV{and DMRG} natural-orbital-based active spaces by diagonalizing the one-particle density matrices. We denote the $n$-electron $n$-orbital active space as ($n$e, $n$o). We then solved within the active space using ground-state DMRG with bond dimension $M=1200$ and then dynamic DMRG with bond dimension $M=800$~\cite{Zhai2021a,Zhai2023a}. To understand the accuracy of this active-space DMRG solver, we also solved the full 40-orbital SIAM problem using DMRG/DDMRG with the same bond dimensions.

\begin{figure*}[hbt]
\centering
\includegraphics[width=\textwidth]{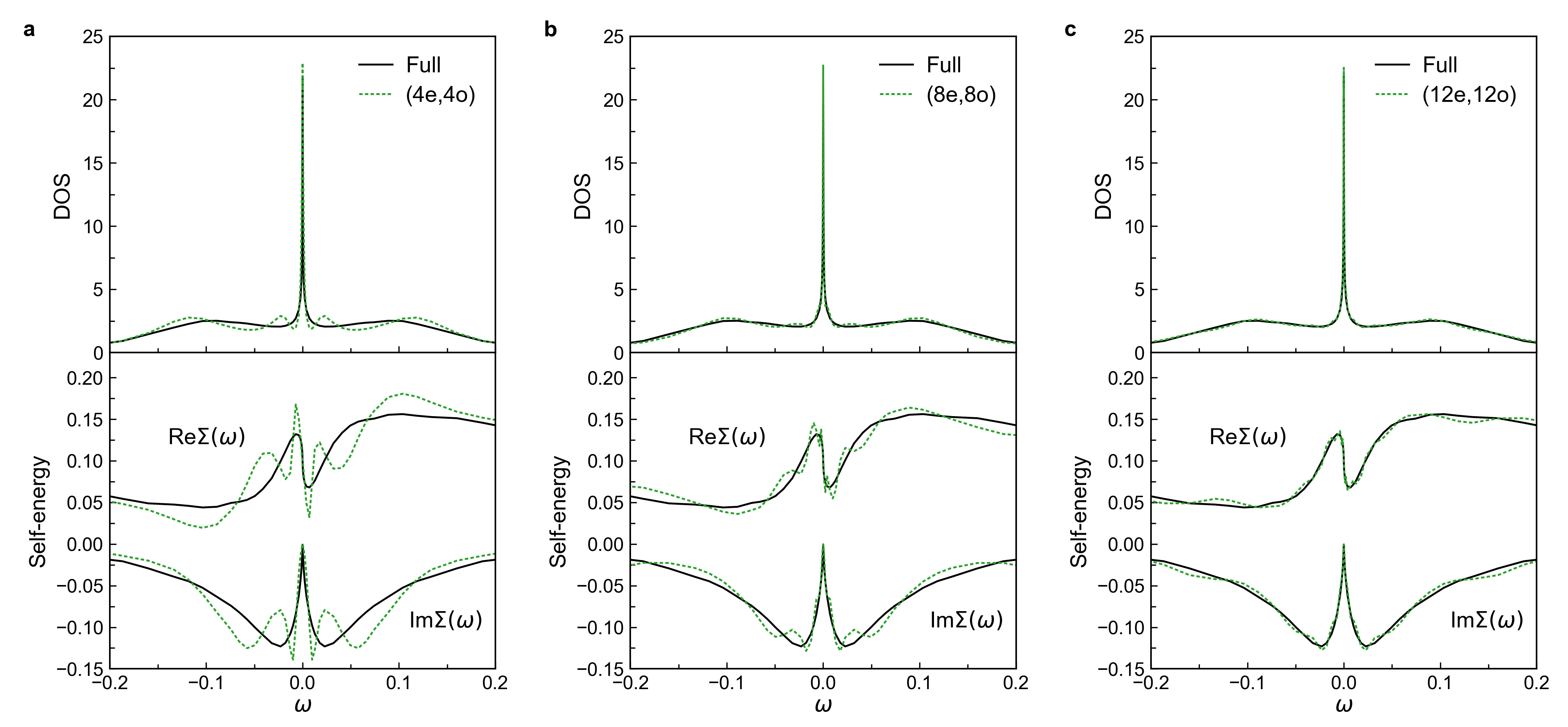}
\caption{Impurity density of states and self-energy in the single-impurity Anderson model calculated by full DMRG and active-space DMRG (in CISD natural orbital basis). $U=0.2$. (a) (4e, 4o) active space. (b) (8e, 8o) active space. (c) (12e, 12o) active space. }
\label{fig:siam}
\end{figure*}

We first present the impurity density of states (DOS) and self-energy at $U=0.2$ in Fig.~\ref{fig:siam}. It shows a sharp Kondo resonance peak and two broad Hubbard bands in the full DMRG calculation, which agrees quantitatively with previous NRG results~\cite{Bulla1998}. Compared to the full (40e, 40o) DMRG results (Fig.~\ref{fig:siam}a), the \REV{CISD} (4e, 4o) active-space result shows multiple spurious oscillations in the DOS and self-energy. However, we point out that even with this very small \REV{CISD} (4e, 4o) active space, the impurity solver correctly predicts the shape of the Kondo resonance around the Fermi level, which is also confirmed in the self-energy comparison around $\omega=0$. We find that the accuracy improves rapidly when the \REV{CISD} active space is increased from (4e, 4o) to (8e, 8o) and (12e, 12o). At the (12e, 12o) level, the DOS and self-energy from active-space DMRG are almost indistinguishable from the full DMRG results.

\begin{figure*}[hbt]
\centering
\includegraphics[width=0.6\textwidth]{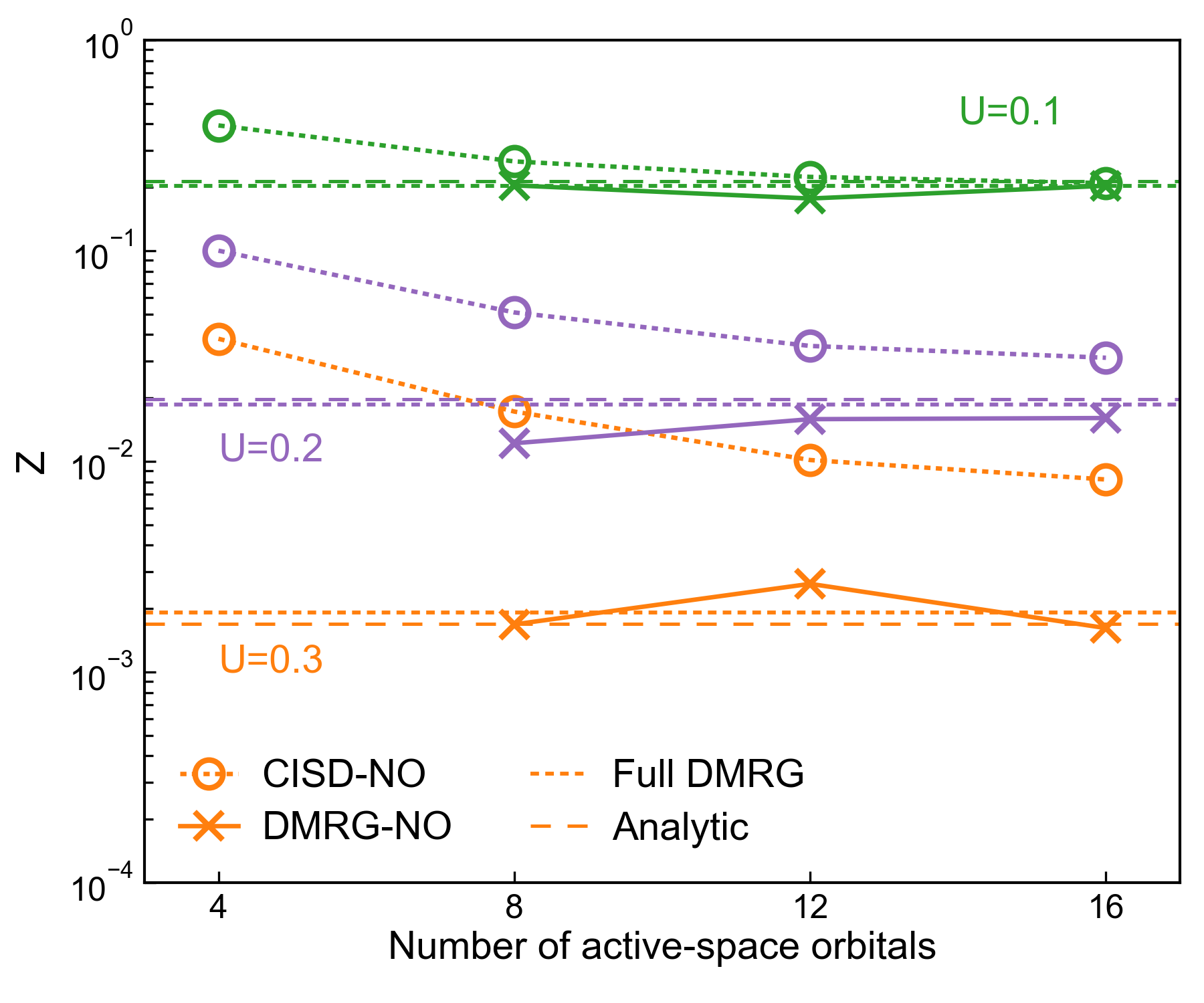}
\caption{Quasiparticle renormalization weight $Z$ for increasing sizes of active space in $U=0.1, 0.2, 0.3$ SIAM models. The active spaces consist of CISD natural orbitals (``CISD-NO'') or DMRG natural orbitals (``DMRG-NO''). The full (40e, 40o) DMRG calculated $Z$ and analytic (exact) $Z$ values are shown as horizontal dashed lines.}
\label{fig:siam2}
\end{figure*}

In Fig.~\ref{fig:siam2}, we further compare the quasiparticle renormalization factors $Z$ defined as
\begin{equation}
    Z = \Big[1 - \frac{\partial \Sigma(\omega)}{\partial \omega} \Big|_{\omega=0} \Big]^{-1}
\end{equation}
\REV{using the CISD and DMRG active spaces, denoted ``CISD-NO'' and ``DMRG-NO,'' respectively.} 
At moderate interaction strength ($U=0.1$), the active-space DMRG predicts $Z=0.21$ at the CISD (16e, 16o) level, in excellent agreement with full DMRG $Z=0.20$ and the exact $Z=0.21$~\cite{Hewson1993a}. The exact $Z$ is analytically calculated from Hewson's formula $Z=4T_\mathrm{K}/\pi \Delta_0$ (with $\Delta_0=0.015$) and the $T_\mathrm{K}$ for large $\frac{U}{\Delta_0}$~\cite{Hewson1993a}
\begin{equation}
    T_\mathrm{K}=\sqrt{\frac{U\Delta_0}{2}} \exp \left( - \frac{\pi U}{8\Delta_0} + \frac{\pi \Delta_0}{2 U}\right).
\end{equation}

However, at stronger interaction strengths, the CISD active spaces become less accurate and overestimate $Z$. In contrast, the DMRG active spaces are more accurate, and 
even with the small (8e, 8o) active space, the computed $Z$ values are reasonably close to the full DMRG and exact results: at $U=0.2$ ($U=0.3$), the active-space DMRG predicts $Z=0.012$ ($Z=0.0017$), while the full DMRG predicts $Z=0.019$ ($Z=0.0019$) and the exact result is $Z=0.020$ ($Z=0.0017$). With the DMRG (16e, 16o) active space, the computed $Z$ values are almost indistinguishable from the exact results on the scale of Fig.~\ref{fig:siam2}.
This confirms the effectiveness of an active-space strategy to reduce the number of orbitals needed to represent the properties of the impurity.

\begin{table}[hbt!]
	\centering
	\caption{Orbital occupancy cut-off thresholds for DMRG occupied and virtual natural orbitals to obtain active spaces of 6-16 orbitals in $U=0.2$ SIAM model.}
	\label{tab:asthresh}
	\begin{tabular}{>{\centering\arraybackslash}p{3.5cm}>{\centering\arraybackslash}p{3.5cm}>{\centering\arraybackslash}p{3.5cm}}
	\hline\hline
	Active space size & Threshold (occupied) & Threshold (virtual) \\
	\hline
        6o & $2\times 10^{-3}$ & $2\times 10^{-3}$ \\
        8o & $4\times 10^{-4}$ & $4\times 10^{-4}$ \\
        10o & $9\times 10^{-5}$ & $8\times 10^{-5}$ \\
        16o & $2\times 10^{-6}$ & $8\times 10^{-7}$ \\
    \hline\hline
     \end{tabular}
\end{table}

\begin{table}[hbt!]
	\centering
	\caption{Orbital occupancy cut-off thresholds for occupied and virtual natural orbitals to obtain the largest active spaces used in dynamical and finite temperature calculations (around 28 orbitals) in \textit{ab initio} calculations of various impurity elements.}
	\label{tab:abinitiothresh}
	\begin{tabular}{>{\centering\arraybackslash}p{3.5cm}>{\centering\arraybackslash}p{3.5cm}>{\centering\arraybackslash}p{3.5cm}}
	\hline\hline
	Element & Threshold (occupied) & Threshold (virtual) \\
	\hline
        Ti & $4\times 10^{-5}$  & $4\times 10^{-4}$  \\
        V & $2\times 10^{-4}$  & $7\times 10^{-4}$  \\
        Cr & $4\times 10^{-4}$  & $3\times 10^{-3}$  \\
        Mn & $2\times 10^{-4}$  & $4\times 10^{-3}$  \\
        Fe & $1\times 10^{-4}$  & $4\times 10^{-3}$  \\
        Co & $1\times 10^{-4}$  & $1\times 10^{-3}$  \\
        Ni & $2\times 10^{-5}$  & $2\times 10^{-3}$  \\
    \hline\hline
     \end{tabular}
\end{table}

\REV{We cannot rigorously infer the accuracy of the active-space DMRG in the \textit{ab initio} Kondo simulations from the SIAM calculations because of the different Hamiltonian. However, we can assign a rough correspondence to the SIAM problem by dividing the number of active space orbitals used in the \textit{ab initio} calculation by the number of effective singly occupied orbitals in the impurity atom, and using $Z$ in the \textit{ab initio} calculations to define a corresponding effective $U$. The most difficult systems are those with the lowest Kondo temperatures, namely Cr, Mn, and Fe, which have between 3-5 singly occupied orbitals. For Cr and Fe, the largest active space used in the computation of $Z$ had 28 orbitals, thus we infer an effective SIAM active space size of 6-10 orbitals. Similar correspondence can be obtained by comparing the natural orbital cut-off thresholds used to obtain active spaces in the \textit{ab initio} calculations and in the SIAM model (Tables~\ref{tab:asthresh} and \ref{tab:abinitiothresh}). The cut-off thresholds in \textit{ab initio} calculations are on the order of $10^{-4}$ or $10^{-3}$, which are in the range of the thresholds of SIAM active spaces of 6-10 orbitals. The renormalization factors of Cr and Fe computed in the largest active space were 0.023 and 0.026, respectively. These $Z$ values correspond roughly to  $U=0.2$, where with this effective (CISD) active space size in the SIAM problem we expect $Z$ to be overestimated by $\sim 3$, which is similar to the observed overestimation of $T_\mathrm{K}$ with respect to the experimental estimates in the \textit{ab initio} calculations. For Mn, where we could only use a 22-orbital active space (see below) and $Z \approx 0.006$, corresponding to $U > 0.2 \sim 0.3$, we expect $Z$ to be overestimated by an order of magnitude. This also roughly agrees with the larger overestimation of $T_\mathrm{K}$ for Mn. The convergence of the SIAM calculations can thus be seen to be consistent with our \textit{ab initio} results, and further provide an estimate of the required resources  for more accurate \textit{ab initio} calculations in the future.} 


\section{Convergence of impurity observables towards parent basis limit}

\begin{table}[hbt!]
	\centering
	\caption{Occupancies ($n$) and quasiparticle renormalization weights ($Z$) of impurity $t_{2g}$ and $e_g$ orbitals obtained from different DMRG/DDMRG active-space simulations. All results are in the def2-SVP basis unless specified.}
	\label{tab:active}
	\begin{tabular}{>{\centering\arraybackslash}p{1.7cm}>{\centering\arraybackslash}p{2.3cm}>{\centering\arraybackslash}p{2.3cm}>{\centering\arraybackslash}p{2cm}>{\centering\arraybackslash}p{2cm}>{\centering\arraybackslash}p{2cm}}
	\hline\hline
	Impurity  & Active space & $n(t_{2g})$ &  $n(e_g)$ & $Z(t_{2g})$ & $Z(e_g)$ \\
	\hline
    \textbf{Ti}   &  (10e, 10o)   & 0.59	& 0.26	& 0.64	& 0.59  \\
     & (18e, 18o) & 0.60 &	0.30	& 0.59	& 0.62 \\
     & (22e, 22o) & 0.60 &	0.32	& 0.58	& 0.58 \\
     & (30e, 29o) & 0.61	& 0.30	& 0.60	& 0.59 \\
     & (36e, 52o) & 0.62	& 0.32 &  & \\
     \hline
     \textbf{V}   &  (10e, 10o)   & 0.99	& 0.19	& 0.27	& 0.45 \\
     & (18e, 18o) & 0.93	& 0.37	& 0.20	& 0.47 \\
     & (22e, 22o) & 0.93	& 0.39	& 0.19	& 0.37 \\
     & (28e, 30o) & 0.93	& 0.38	& 0.19	& 0.37 \\
     & (36e, 52o) & 0.95	& 0.41 &  & \\
     \hline
     \textbf{Cr}   &  (10e, 10o)   & 0.91	& 0.82	& 0.12	& 0.12 \\
     & (14e, 16o) & 1.10	& 0.61	& 0.092	& 0.16 \\
     & (22e, 22o) & 0.97 &	0.94	& 0.017	& 0.038 \\
     & (28e, 28o) & 0.99	& 0.93	& 0.023	& 0.047 \\
     & (36e, 52o) & 0.99	& 0.96 &  & \\
     \hline
     \textbf{Mn}   &  (10e, 10o)   & 1.06	& 1.07	& 0.045 &	0.011 \\
     & (16e, 16o) & 1.16	& 1.00	& 0.068	& 0.018 \\
     & (22e, 22o) & 1.17	& 1.03	& 0.071	& 0.0058 \\
     & (36e, 52o) & 1.22	& 1.03 &  & \\
     \hline
     \textbf{Fe}   &  (10e, 10o)   & 1.46	& 1.14	& 0.27	& 0.12 \\
     (def2-SVP) & (16e, 16o) & 1.59 &	1.03	& 0.23	& 0.021 \\
     & (22e, 22o) & 1.58	& 1.06	& 0.15	& 0.028 \\
     & (28e, 28o) & 1.59	& 1.04	& 0.15	& 0.026 \\
     & (36e, 52o) & 1.58	& 1.08 &  & \\
     \hline
     \textbf{Co}   &  (10e, 10o)   & 1.85	& 1.16	& 0.29	& 0.11 \\
     & (16e, 16o) & 1.90	& 1.09	& 0.37	& 0.060 \\
     & (22e, 27o) & 1.80 &	1.26	& 0.19	& 0.11 \\
     & (26e, 36o) & 1.79	& 1.28	& 0.20	& 0.11 \\
     & (36e, 52o) & 1.77	& 1.32 &  & \\
     \hline
     \textbf{Ni}   &  (10e, 10o)   & 1.77	& 1.77	& 0.50	& 0.47 \\
     & (16e, 16o) & 1.74	& 1.85	& 0.48	& 0.47 \\
     & (22e, 22o) & 1.80	& 1.80	& 0.42	& 0.40 \\
     & (28e, 28o) & 1.78	& 1.81	& 0.40	& 0.38 \\
     & (36e, 46o) & 1.80	& 1.82 &  & \\
     \hline
     \textbf{Fe}   &  (26e, 29o)   &  1.54	& 1.08	& 0.19	& 0.039 \\
     (cc-pVTZ) & (46e, 76o) & 1.56	& 1.08 & & \\
    \hline\hline
     \end{tabular}
\end{table}

\begin{figure*}[hbt!]
\centering
\includegraphics{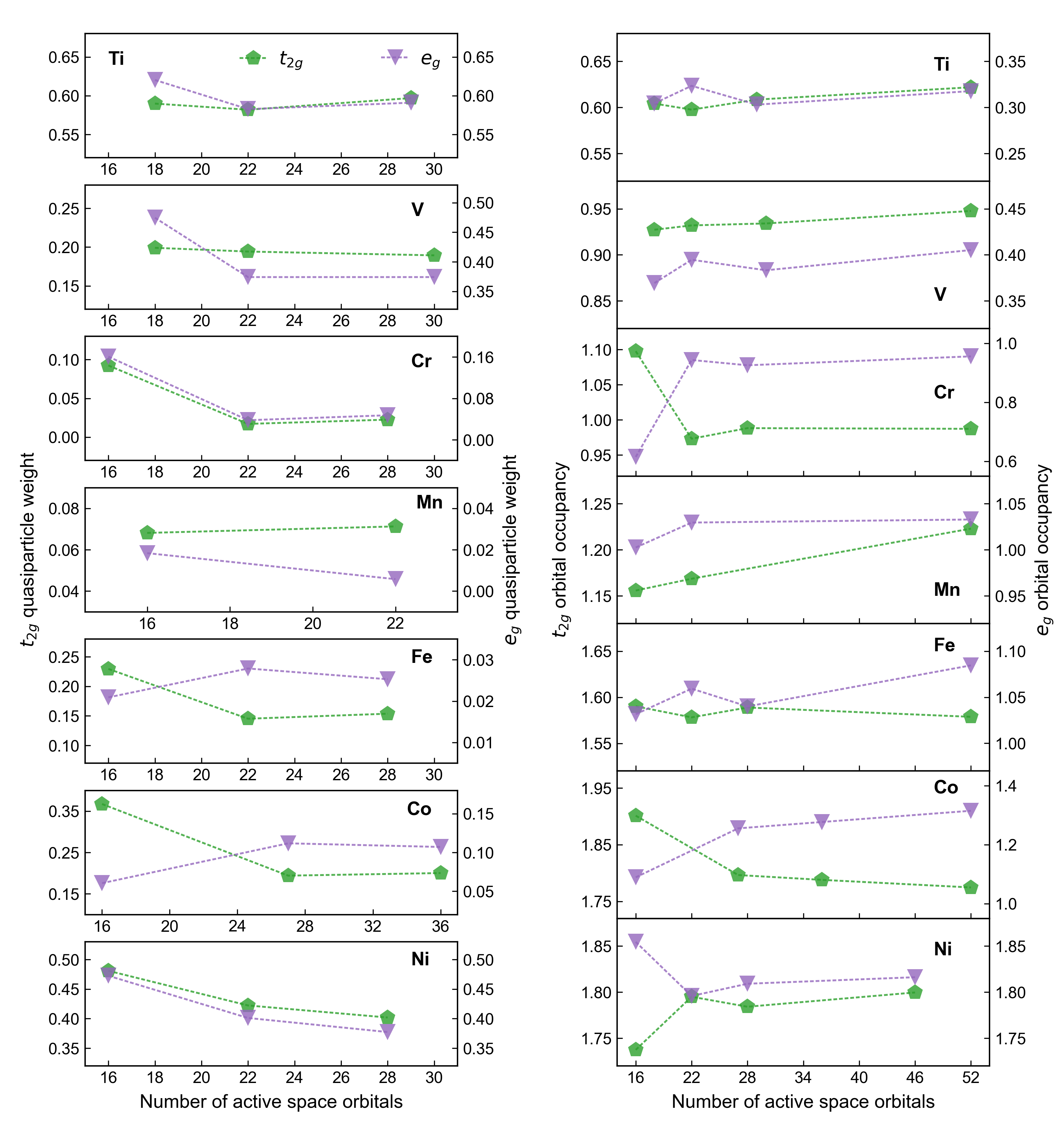}
\caption{Convergence of the quasiparticle renormalization weights (left) and orbital occupancies (right) of the $3d$ orbitals of all impurities as the DMRG active space increases.}
\label{fig:activespacesi}
\end{figure*}

In this section, we present detailed benchmarks of the convergence of local impurity observables in the active-space DMRG calculations towards the parent basis limit. By diagonalizing the one-particle density matrix obtained from ground-state DMRG calculations within the (36e, 52o) active space ((36e, 46o) for Ni) in the def2-SVP basis, we further derived a series of smaller active spaces for performing the more time-consuming dynamical DMRG (DDMRG) calculations. The tested active spaces are listed in Table~\ref{tab:active}, where the corresponding orbital occupancies $n$ and quasiparticle (QP) renormalization weights $Z$ of impurity $3d$ orbitals are shown in Table~\ref{tab:active} and Fig.~\ref{fig:activespacesi}. We find that, with the exception of Mn (where we could not converge the DDMRG calculation with respect to bond dimension for an active space larger than (22e, 22o)), $Z$ and $n$ are converged with small errors for all impurities at an active-space size of (22e, 22o) ((22e, 27o) for Co).

\clearpage
\newpage

\section{Many-body solver error estimates}
We next present an estimate of the errors from two of the approximations in our calculations: active space size and DMRG bond dimension, which are the main many-body solver errors with respect to the exact parent-basis limit (in this case, def2-SVP). First, we analyze the errors in the predicted impurity observables due to the use of active spaces. As shown in Table~\ref{tab:aserror}, for orbital occupancies, we estimate the remaining errors relative to the parent-basis limit by comparing the largest and second largest ground-state (GS) active-space calculations (assuming (36e, 52o) is a sufficiently large active space). For example, for Fe, we obtain $\Delta n(t_{2g})=0.010$ and $\Delta n(e_{g})=0.045$ by comparing the 52o (largest GS) and 28o (second largest GS) results. Using the maximum $\Delta n$ across all impurities (except Mn), we estimate that the active space associated orbital occupancy error to be at most 0.045. For estimating the QP weight errors, we adopt a two-step procedure: (1) Compare the largest and second largest Green's function (GF) active-space calculations. For Fe, we obtain $\Delta Z(t_{2g})=0.009$ and $\Delta n(e_{g})=0.002$ by taking the difference between the 28o (largest GF) and 22o (second largest GF) results. (2) Estimate the QP weight errors between the largest and second largest GS active spaces by assuming a linear relationship between $\Delta n$ and $\Delta Z$:
\begin{equation}
    \Delta Z(\text{largest GS vs. 2nd largest GS}) = \frac{\Delta n(\text{largest GS vs. 2nd largest GS})}{\Delta n(\text{largest GF vs. 2nd largest GF})} \times \Delta Z(\text{largest GF vs. 2nd largest GF}).
    \label{eq:est}
\end{equation}
Following this procedure, we estimate the $Z$ errors for Fe to be $\Delta Z(t_{2g})=0.008$ and $\Delta n(e_{g})=0.005$. Furthermore, using the maximum $\Delta Z$ across all impurities (except Mn), we estimate the active space associated QP weight error to be at most 0.048.

\begin{table}[hbt!]
	\centering
	\caption{Active-space error analysis of orbital occupancies ($n$) and quasiparticle renormalization weights ($Z$) of impurity $3d$ impurity orbitals. The ``est.'' values are estimated according to Eq.~\ref{eq:est}.}
	\label{tab:aserror}
	\begin{tabular}{>{\centering\arraybackslash}p{1.8cm}>{\centering\arraybackslash}p{3cm}>{\centering\arraybackslash}p{1.8cm}>{\centering\arraybackslash}p{1.8cm}>{\centering\arraybackslash}p{2cm}>{\centering\arraybackslash}p{2cm}}
	\hline\hline
	Impurity  & Comparison btw. active spaces & $\Delta n(t_{2g})$ & $\Delta n(e_{g})$ & $\Delta Z(t_{2g})$ & $\Delta Z(e_g)$ \\
	\hline
    \textbf{Ti}   &  22o vs. 29o  & 	0.011 &	0.020	& 0.015	& 0.008 \\
    &  29o vs. 52o  & 	0.013 &	0.015 &	0.018 (est.)	& 0.006 (est.) \\
    \hline
    \textbf{V}   &  22o vs. 30o  & 	0.002	& 0.011	& 0.007	& 0.001 \\
    &  30o vs. 52o  & 	0.014	& 0.022 &	0.048 (est.)	& 0.001 (est.) \\
    \hline
    \textbf{Cr}   &  22o vs. 28o  & 	0.015	& 0.018	& 0.005	& 0.010 \\
    &  28o vs. 52o  & 	0.001	& 0.030 &	0.000 (est.)	& 0.016 (est.) \\
    \hline
    \textbf{Fe}   &  22o vs. 28o  & 	0.011	& 0.019	& 0.009	& 0.002 \\
    &  28o vs. 52o  & 	0.010	& 0.045 &	0.008 (est.)	& 0.005 (est.) \\
    \hline
    \textbf{Co}   &  22o vs. 36o  & 	0.009	& 0.022	& 0.006	& 0.005 \\
    &  36o vs. 52o  & 	0.013	& 0.038 &	0.010 (est.)	& 0.008 (est.) \\
    \hline
    \textbf{Ni}   &  22o vs. 28o  & 	0.011	& 0.013	& 0.020	& 0.024 \\
    &  28o vs. 46o  & 	0.015	& 0.007 &	0.029 (est.)	& 0.013 (est.) \\
    \hline\hline
     \end{tabular}
\end{table}

\begin{table}[hbt!]
	\centering
	\caption{Error analysis of quasiparticle renormalization weights of impurity $3d$ orbitals against DDMRG maximum bond dimensions and discarded weights. The ``extrap.'' values are obtained through two-point linear extrapolation against discarded weights.}
	\label{tab:gfdmrgbond}
	\begin{tabular}{>{\centering\arraybackslash}p{2cm}>{\centering\arraybackslash}p{4cm}>{\centering\arraybackslash}p{3.5cm}>{\centering\arraybackslash}p{1.8cm}>{\centering\arraybackslash}p{1.8cm}}
	\hline\hline
	Impurity  & Max.~bond dimension & Max.~discarded weight & $Z(t_{2g})$ & $Z(e_g)$ \\
	\hline
    \textbf{V}   &  2500  & $1.52 \times 10^{-2}$	& 0.1893 & 0.3742	\\
    (28e, 30o) & 4000 & $9.41 \times 10^{-3}$ & 0.1867 & 0.3736 \\
    & extrap. & 0 & 0.1825 & 0.3726 \\
    & $|$$Z$($M=4000$)$-$$Z$(extrap.)$|$ &  & $4.2 \times 10^{-3}$ & $1.0 \times 10^{-3}$ \\
    \hline
    \textbf{Cr}   &  2500  & $2.24 \times 10^{-2}$	& 0.0202	& 0.0473	\\
    (28e, 28o) & 3500 & $1.47 \times 10^{-2}$ & 0.0226	& 0.0474 \\
    & extrap. & 0 & 0.0273 & 0.0476 \\
    & $|$$Z$($M=3500$)$-$$Z$(extrap.)$|$ &  & $4.7 \times 10^{-3}$ & $1.9 \times 10^{-4}$ \\
    \hline
    \textbf{Fe}   &  2500  & $5.28 \times 10^{-3}$	& 0.1538	& 0.0253	\\
    (28e, 28o) & 3500 & $2.97 \times 10^{-3}$ & 0.1538	& 0.0255 \\
    & extrap. & 0 & 0.1539 & 0.0258 \\
    & $|$$Z$($M=3500$)$-$$Z$(extrap.)$|$ &  & $7.3 \times 10^{-5}$ & $2.9 \times 10^{-4}$ \\
    \hline
    \textbf{Co}   &  1500  & $3.46 \times 10^{-2}$	& 0.2011	& 0.1088	\\
    (26e, 36o) & 2500 & $1.29 \times 10^{-3}$ & 0.2003	& 0.1071 \\
    & extrap. & 0 & 0.1999 & 0.1062 \\
    & $|$$Z$($M=2500$)$-$$Z$(extrap.)$|$ &  & $4.5 \times 10^{-4}$ & $9.9 \times 10^{-4}$ \\
    \hline\hline
     \end{tabular}
\end{table}

We then analyze the numerical errors due to finite bond dimension in DMRG calculations. We find that the DMRG predicted orbital occupancies are very well converged with respect to the bond dimension ($M$) in all impurities. For example, for Cr, the orbital occupancy differences between $M=3000$ and $M=4000$ calculations are negligibly small: $\Delta n(t_{2g})=2\times 10^{-4}$ and $\Delta n(e_{g})=1\times 10^{-4}$. 

We then focus on the DMRG bond dimension associated QP weight errors. In Table~\ref{tab:gfdmrgbond}, we present DDMRG $Z$ values and discarded weights at various bond dimensions for V, Cr, Fe, and Co, in their largest GF active-space calculations. We also performed a two-point linear extrapolation against the discarded weights to estimate the $Z$ values at infinite DDMRG bond dimension. By comparing the largest $M$ results against the extrapolated values, we find that the largest bond dimension associated $Z$ error is around $5 \times 10^{-3}$ (Cr), an order of magnitude smaller than the largest active space associated $Z$ error. 

In summary, combining the maximum errors from the active space and bond dimension error analysis, we conservatively estimate that our predicted $n$ values are converged to at least $\sim$ 0.04 and $Z$ values are converged to at least $\sim$ 0.05 compared to the exact parent basis result.

\section{DFT hybridization functions}
\begin{figure*}[hbt!]
\centering
\includegraphics[width=\textwidth]{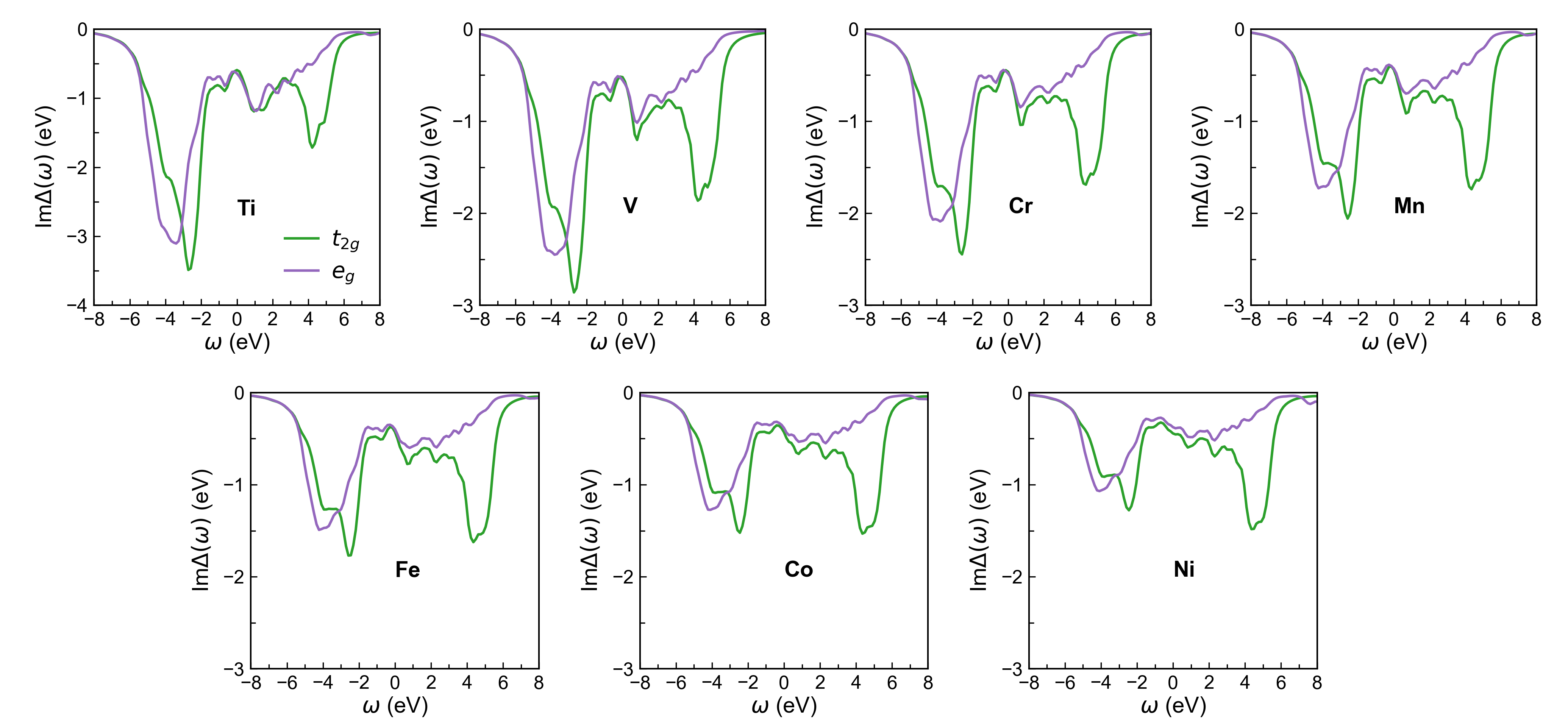}
\caption{Real-axis hybridization function of $3d$ orbitals of magnetic impurities in bulk Cu calculated by DFT with the PBE functional.}
\label{fig:hyb}
\end{figure*}

We present DFT-calculated real-axis hybridization functions of magnetic impurity atoms in bulk Cu in Fig.~\ref{fig:hyb}. It is observed that the $t_{2g}$ hybridization has a greater magnitude than the $e_{g}$ hybridization in all impurities, especially in the range of [3, 6] eV. Meanwhile, we find that the magnitudes of both $t_{2g}$ and $e_g$ hybridization functions become smaller from Ti to Ni. 

\begin{figure*}[hbt!]
\centering
\includegraphics[width=0.68\textwidth]{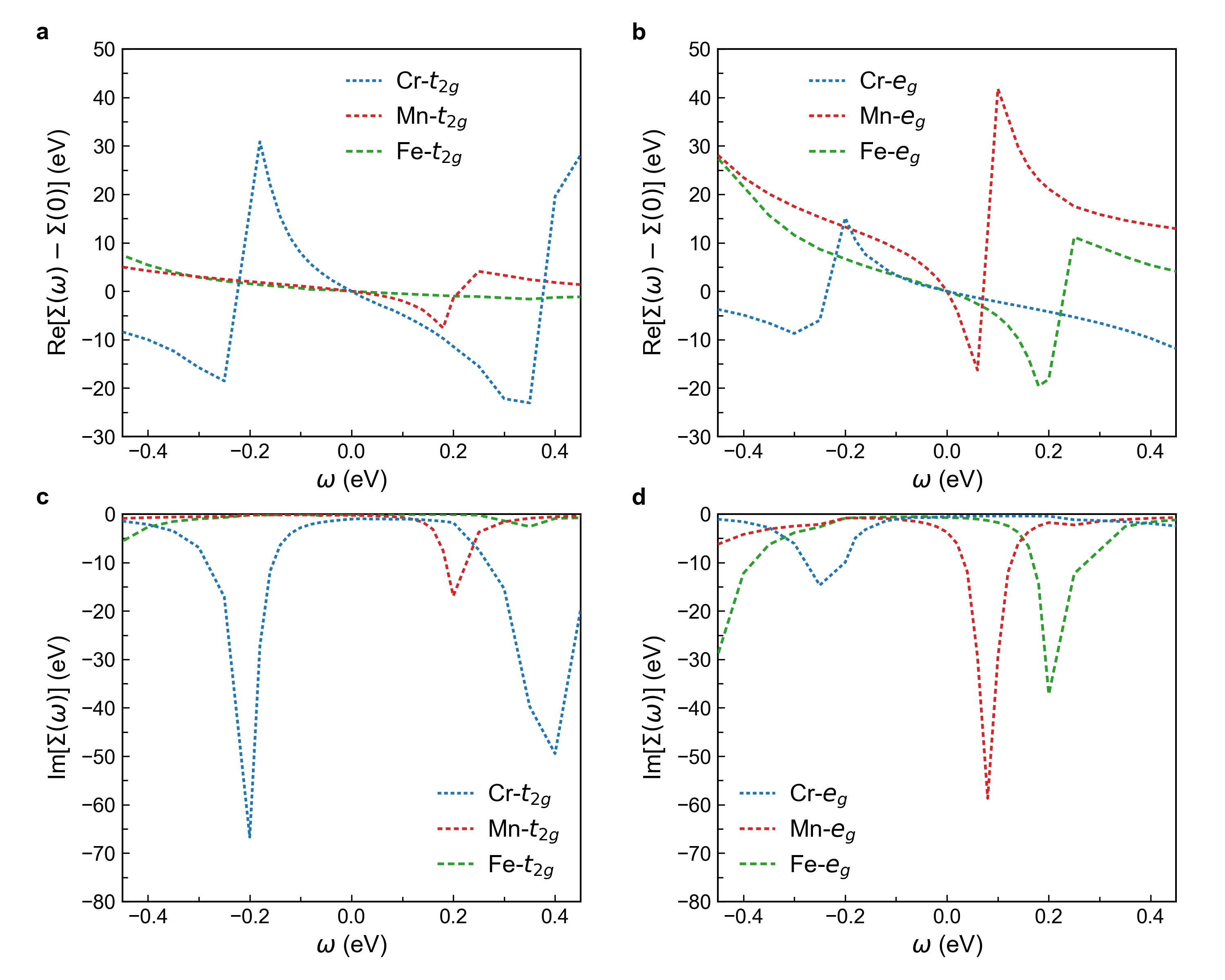}
\caption{Real-axis self-energies of Cr, Mn, and Fe impurities from all-orbital calculations. (a) Real part of \REV{the self-energy matrix elements diagonal in the orbital indices associated with the }$t_{2g}$ orbitals. (b) Real part of \REV{the self-energy matrix elements diagonal in the orbital indices associated with the }$e_{g}$ orbitals. (c) Imaginary part of self-energies of $t_{2g}$ orbitals. (d) Imaginary part of self-energies of $e_{g}$ orbitals.}
\label{fig:selfenergy}
\end{figure*}

\section{Self-energies of magnetic impurity orbitals}
We show real-axis self-energies of Cr, Mn, and Fe impurities in bulk Cu calculated by all-orbital simulations in Fig.~\ref{fig:selfenergy}. When $\omega$ approaches the Fermi level, the imaginary part of the self-energies of the $3d$ orbitals approaches zero, while the real part of the self-energies of the $3d$ orbitals changes linearly with respect to the frequency, which agrees with the expected Fermi liquid behavior.

\begin{figure*}[hbt!]
\centering
\includegraphics[width=0.68\textwidth]{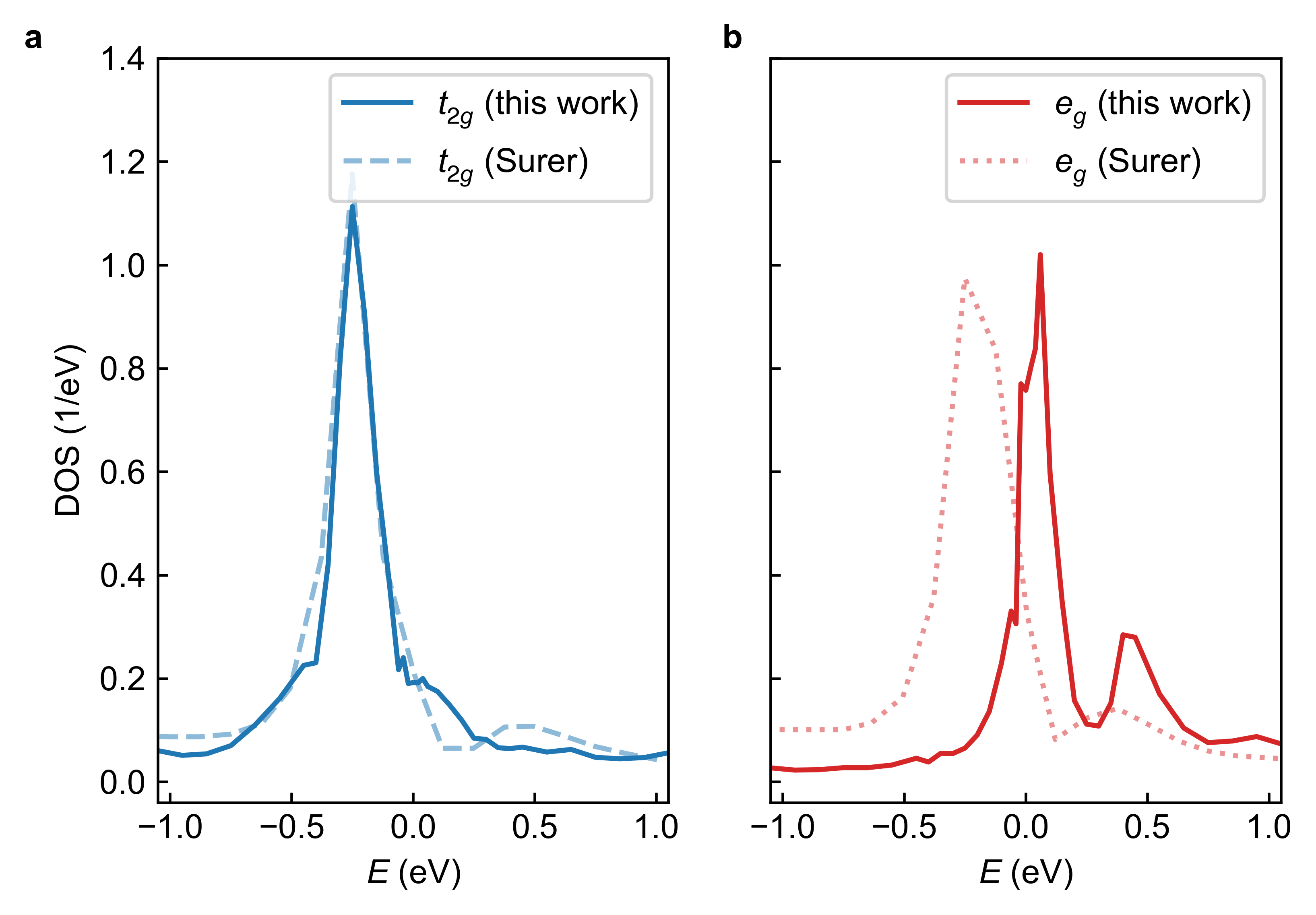}
\caption{\REV{(a) $t_{2g}$ and (b) $e_g$ resolved DOS of Co impurity in bulk Cu from all-orbital calculations (``this work''), compared to DOS taken from Ref.~\cite{Surer2012a} (``Surer'').}}
\label{fig:Co_dos}
\end{figure*}

\section{Density of state of magnetic impurity orbitals}
\REV{We present the local DOS of the $3d$ orbitals of the Co impurity in bulk Cu calculated by all-orbital simulations in Fig.~\ref{fig:Co_dos}, in addition to the DOS of Cr, Mn and Fe impurities in Figure 2. Compared to results from the 5-orbital model calculations in Ref.~\cite{Surer2012a}, the all-orbital DOS exhibits qualitatively similar resonance and shoulder peaks near the Fermi level. Quantitatively, resonance peaks near the Fermi level from all-orbital calculations are narrower, especially for the $e_g$ orbitals. The narrower widths are consistent with the smaller quasiparticle renormalization weights $Z$ = 0.19 ($t_{2g}$) and 0.11 ($e_{g}$) from our all-orbital calculation compared to the $Z$ = 0.42 ($t_{2g}$) and 0.47 ($e_{g}$) from Ref.~\cite{Surer2012a}. }

\section{Summary of ground-state and spectral properties}

The ground-state and spectral properties of magnetic impurities calculated by all-orbital simulations are summarized in Table~\ref{tab:tk} and Table~\ref{tab:density}. In Table~\ref{tab:density}, we include the natural orbital occupancies of $t_{2g}$ and $e_g$ symmetry derived from the (36e, 52o) active-space DMRG density matrix. We choose to show the most fractional occupancies with $n_\mathrm{nat} > 1$. We observe that the natural occupancies in Table~\ref{tab:density} correlate well with the quasiparticle renormalization $Z$ in Table~\ref{tab:tk}, i.e., more fractionally occupied orbitals are found to possess stronger quasiparticle renormalization.

\begin{table}[hbt!]
	\centering
	\caption{Hybridization function values at the Fermi level $\Delta(0)$, quasiparticle renormalization weights $Z$, and Kondo temperatures $T_\mathrm{K}$ of magnetic impurities in bulk Cu computed by all-orbital simulations. All results are in the def2-SVP basis with Green's function active space of (22e, 22o) ((22e, 27o) for Co).}
	\label{tab:tk}
	\begin{tabular}{>{\centering\arraybackslash}p{1.7cm}>{\centering\arraybackslash}p{2.3cm}>{\centering\arraybackslash}p{2.3cm}>{\centering\arraybackslash}p{1.7cm}>{\centering\arraybackslash}p{1.7cm}>{\centering\arraybackslash}p{1.9cm}>{\centering\arraybackslash}p{1.9cm}}
	\hline\hline
	Impurity  & $\Delta(0, t_{2g})$ (eV) & $\Delta(0, e_g)$ (eV) & $Z(t_{2g})$ & $Z(e_{g})$ & $T_\mathrm{K}(t_{2g})$ (K) & $T_\mathrm{K}(e_{g})$ (K) \\
	\hline
    Ti   &  $-0.59$   & $-0.64$  & 0.58 & 0.58  & 3144 & 3412 \\
    V & $-0.52$ & $-0.56$ & 0.19 & 0.37 & 924 & 1895 \\
    Cr & $-0.47$ & $-0.48$ & 0.017 & 0.038 & 73 & 167 \\
    Mn & $-0.44$ & $-0.43$ & 0.071 & 0.0058 & 287 & 23 \\
    Fe & $-0.43$ & $-0.40$ & 0.15 & 0.028 & 565 & 101 \\
    Co & $-0.43$ & $-0.38$ & 0.19 & 0.11 & 761 & 396 \\
    Ni & $-0.44$ & $-0.38$ & 0.42 & 0.40 & 1709 & 1374 \\
    \hline\hline
     \end{tabular}
\end{table}

\begin{table}[hbt!]
	\centering
	\caption{Orbital occupancies $n$, natural orbital occupancies $n_\mathrm{nat}$, and spin moments $S$ of magnetic impurities in bulk Cu computed by all-orbital simulations. All results are in the def2-SVP basis with ground-state active space of (36e, 52o) ((36e, 46o) for Ni).}
	\label{tab:density}
	\begin{tabular}{>{\centering\arraybackslash}p{2cm}>{\centering\arraybackslash}p{1.8cm}>{\centering\arraybackslash}p{1.8cm}>{\centering\arraybackslash}p{2cm}>{\centering\arraybackslash}p{2cm}>{\centering\arraybackslash}p{1.8cm}>{\centering\arraybackslash}p{1.8cm}}
	\hline\hline
	Impurity  & $n(t_{2g})$ & $n(e_g)$ & $n_\mathrm{nat}(t_{2g})$ & $n_\mathrm{nat}(e_g)$ & $n(\mathrm{3d})$ & $S$ \\
	\hline
    Ti   &  0.62 & 0.32 & 1.84 & 1.93 & 2.50 & 0.93 \\
    V & 0.95 & 0.40 & 1.57 & 1.86 & 3.65 & 1.36 \\
    Cr & 0.99 & 0.96 & 1.24 & 1.21 & 4.87 & 2.23 \\
    Mn & 1.22 & 1.03 & 1.46 & 1.20 & 5.73 & 1.98 \\
    Fe & 1.58 & 1.09 & 1.64 & 1.29 & 6.92 & 1.42 \\
    Co & 1.77 & 1.31 & 1.76 & 1.52 & 7.94 & 0.95 \\
    Ni & 1.80 & 1.82 & 1.89 & 1.89 & 9.03 & 0.48 \\
    \hline\hline
     \end{tabular}
\end{table}

\section{Spin and charge fluctuations in Kondo impurities}
\begin{figure*}[hbt!]
\centering
\includegraphics[width=\textwidth]{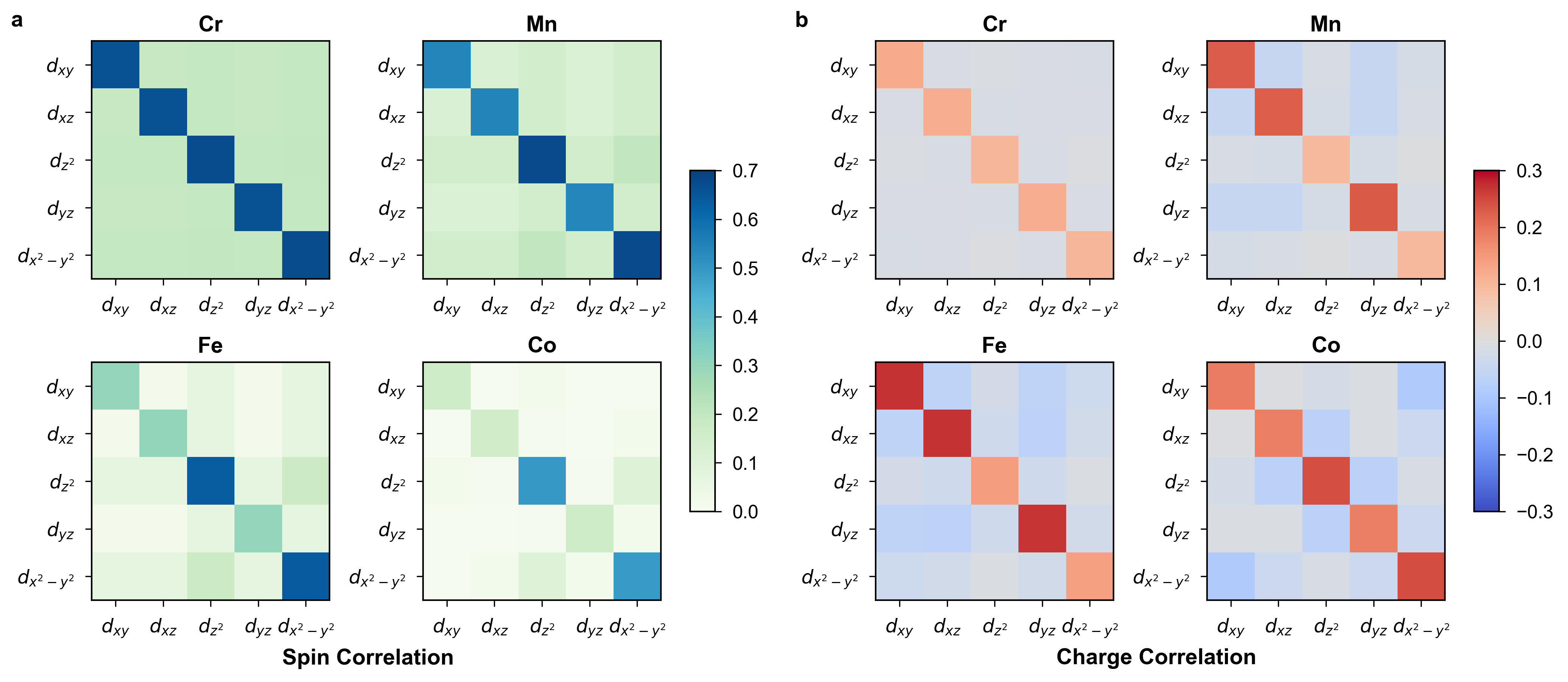}
\caption{Orbital-resolved \REV{(a) spin and (b) charge correlations} for Cr, Mn, Fe, and Co impurities.}
\label{fig:spincorr}
\end{figure*}

The spin fluctuation in all magnetic impurities can be better understood by calculating the spin-spin correlation within the $3d$ shell
\begin{equation}
    S_{ij} = \langle \hat{S}_i \hat{S}_j \rangle - \langle \hat{S}_i \rangle \langle \hat{S}_j \rangle .
\end{equation}
As shown in Fig.~\ref{fig:spincorr}a, Cr has the strongest intra- and inter-orbital spin correlation among all impurities. From Cr to Co, the intra-orbital spin correlation in $t_{2g}$ orbitals is largely suppressed, while the intra- and inter-orbital spin correlation within the $e_g$ orbitals stays strong or moderate, which is consistent with the orbital-dependent trend in quasiparticle renormalization. Fig.~\ref{fig:spincorr}a also reveals that, in the Mn, Fe, and Co systems, the inter-orbital spin correlations have the following relation: $S_{i \neq j} (t_{2g} \text{-} t_{2g}) < S_{i \neq j} (t_{2g} \text{-} e_{g}) < S_{i \neq j} (e_{g} \text{-} e_{g})$. We also include the impurity charge fluctuation in Fig.~\ref{fig:spincorr}b, which is calculated as $C_{ij} = \langle \hat{n}_i \hat{n}_j \rangle - \langle \hat{n}_i \rangle \langle \hat{n}_j \rangle$.

\section{Effective $3d$-model calculations}
To compare our all-orbital simulations against downfolded effective model calculations, we derived a multi-orbital Anderson impurity model with Hamiltonian
\begin{equation}
    H = \sum_{ij\sigma} \tilde{F}_{ij} f^\dagger_{i\sigma} f_{j\sigma} + \frac{1}{2} \sum_{ijkl} \sum_{\sigma \sigma'} U_{ijkl} f^\dagger_{i\sigma} f^\dagger_{j\sigma'} f_{l\sigma'} f_{k\sigma} + \sum_{k \sigma} \epsilon_k c^\dagger_{k\sigma} c_{k\sigma} + \sum_{ki\sigma} V_{ik} (f^\dagger_{i\sigma} c_{k\sigma} + c^\dagger_{k\sigma} f_{i\sigma}),
    \label{eq:MAIM}
\end{equation}
where the indices $i,j,k,l$ run over five $3d$ orbitals and $U_{ijkl}$ is the screened Coulomb interaction tensor within the $3d$ shell. In practice, approximations such as the density-density approximation are often employed for the Coulomb interaction tensor. Here, we adopted a Kanamori Hamiltonian~\cite{Valli2020} that goes beyond the density-density approximation, where the Coulomb tensor (second term in Eq.~\ref{eq:MAIM}) is written as
\begin{equation}
    H_K = \sum_i U_{iiii} \hat{n}_{i\uparrow} \hat{n}_{i\downarrow} + \sum_{i \neq j} \sum_{\sigma \sigma'} (U_{ijij}-U_{ijji} \delta_{\sigma \sigma'}) \hat{n}_{i\sigma} \hat{n}_{j\sigma'} + \sum_{i \neq j} U_{ijji} (f^\dagger_{i\uparrow} f^\dagger_{j\downarrow} f_{i\downarrow} f_{j\uparrow} - f^\dagger_{i\uparrow} f^\dagger_{i\downarrow} f_{j\uparrow} f_{j\downarrow}).
    \label{eq:Kanamori}
\end{equation}
We give the Coulomb integrals for $3d$ electrons in the basis of cubic harmonics, where $U_{iiii}$, $U_{ijij}$, and $U_{ijji}$ are expressed using parameters $U_0, J_1, J_2, J_3, J_4$. The readers are referred to Ref.~\cite{Valli2020} for the detailed parametrization. The $U_0, J_1, J_2, J_3, J_4$ parameters can be expressed in terms of Slater integrals $F^0$, $F^2$, and $F^4$:
\begin{align}
    U_0 &= F^0 + \frac{8}{7} \frac{1}{14} (F^2+F^4), \\
    J_1 &= \frac{1}{49} (3F^2 + \frac{20}{9} F^4) , \\
    J_2 &= -2 \frac{5}{7} \frac{1}{14} (F^2 + F^4) + 3 J_1, \\
    J_3 &= 6 \frac{5}{7} \frac{1}{14} (F^2 + F^4) - 5J_1, \\
    J_4 &= 4 \frac{5}{7} \frac{1}{14} (F^2 + F^4) - 3 J_1.
\end{align}
The Slater integrals $F^0$, $F^2$, and $F^4$ are obtained from two parameters $U$ and $J$, where $U=F_0$ and $J=\frac{1}{14} (F^2+F^4)$ with a constant ratio $F^2/F^4=0.625$. In summary, the Kanamori Hamiltonian in Eq.~\ref{eq:Kanamori} is fully characterized by two parameters $U$ and $J$. In Kondo simulation literature, $U$ and $J$ are usually treated as adjustable parameters. Here, we took the $U$ and $J$ values of magnetic impurities from cRPA calculations in Ref.~\cite{Sasioglu2011}, which are listed in Table~\ref{tab:UJ}. 

\begin{table}[hbt!]
	\centering
        \caption{Coulomb interaction parameters in Kanamori Hamiltonian taken from Ref.~\cite{Sasioglu2011}.}
	\label{tab:UJ}
	\begin{tabular}{>{\centering\arraybackslash}p{2.4cm}>{\centering\arraybackslash}p{1.4cm}>{\centering\arraybackslash}p{1.4cm}>{\centering\arraybackslash}p{1.4cm}>{\centering\arraybackslash}p{1.4cm}>{\centering\arraybackslash}p{1.4cm}>{\centering\arraybackslash}p{1.4cm}>{\centering\arraybackslash}p{1.4cm}}
	\hline\hline
	Parameter (eV)  &  Ti & V & Cr & Mn & Fe & Co & Ni \\
	\hline
    $U$   & 3.1 & 3.2 & 4.4 & 4.4 & 3.8 & 4.3 & 3.8  \\
    $J$  & 0.5 & 0.6 & 0.7 & 0.7 & 0.7 & 0.8 & 0.8 \\
    \hline\hline
     \end{tabular}
\end{table}

\begin{table}[hbt!]
	\centering
	\caption{Quasiparticle renormalization weights $Z$ and Kondo temperatures $T_\mathrm{K}$ of magnetic impurities in bulk Cu from five-orbital model Hamiltonian calculations.}
	\label{tab:tkmodel}
	\begin{tabular}{>{\centering\arraybackslash}p{2cm}>{\centering\arraybackslash}p{2cm}>{\centering\arraybackslash}p{2cm}>{\centering\arraybackslash}p{2cm}>{\centering\arraybackslash}p{2cm}}
	\hline\hline
	Impurity  &  $Z(t_{2g})$ & $Z(e_{g})$ & $T_\mathrm{K}(t_{2g})$ (K) & $T_\mathrm{K}(e_{g})$ (K) \\
	\hline
    Ti   &  0.92 & 0.95  & 4953 & 5556 \\
    V &  0.63 & 0.87 & 2993 & 4431 \\
    Cr &  0.31 & 0.55 & 1318 & 2445 \\
    Mn &  0.11 & 0.062 & 449 & 243 \\
    Fe &  0.45 & 0.11 & 1769 & 398 \\
    Co &  0.71 & 0.72 & 2800 & 2524 \\
    Ni &  0.99 & 0.99 & 3999 & 3383 \\
    \hline\hline
     \end{tabular}
\end{table}

\begin{table}[hbt!]
	\centering
	\caption{Orbital occupancies $n$ and natural orbital occupancies $n_\mathrm{nat}$ of magnetic impurities in bulk Cu from five-orbital model Hamiltonian calculations.}
	\label{tab:densitymodel}
	\begin{tabular}{>{\centering\arraybackslash}p{2cm}>{\centering\arraybackslash}p{2cm}>{\centering\arraybackslash}p{2cm}>{\centering\arraybackslash}p{2cm}>{\centering\arraybackslash}p{2cm}>{\centering\arraybackslash}p{2cm}}
	\hline\hline
	Impurity  & $n(t_{2g})$ & $n(e_g)$ & $n_\mathrm{nat}(t_{2g})$ & $n_\mathrm{nat}(e_g)$ & $n(\mathrm{3d})$ \\
	\hline
    Ti   &  0.32 & 0.13 & 1.98 & 1.94 & 1.23  \\
    V & 0.75 & 0.15 & 1.86 & 1.89 & 2.54  \\
    Cr & 1.23 & 0.27 & 1.63 & 1.85 & 4.23  \\
    Mn & 1.26 & 0.84 & 1.59 & 1.50 & 5.45  \\
    Fe & 1.60 & 1.07 & 1.84 & 1.52 & 6.95  \\
    Co & 1.68 & 1.74 & 1.93 & 1.80 & 8.51  \\
    Ni & 1.94 & 1.97 & 1.91 & 1.88 & 9.76  \\
    \hline\hline
     \end{tabular}
\end{table}

\begin{table}[hbt!]
	\centering
	\caption{\REV{Five-orbital model Hamiltonian results after fitting the chemical potential. Shown in the table are the changes in the chemical potential $\Delta\mu$, orbital occupancies $n$, quasiparticle renormalization weights $Z$, and Kondo temperature $T_\mathrm{K}$.}}
	\label{tab:fittedmodel}
	\begin{tabular}{>{\centering\arraybackslash}p{1.3cm}>{\centering\arraybackslash}p{1.3cm}>{\centering\arraybackslash}p{1.3cm}>{\centering\arraybackslash}p{1.3cm}>{\centering\arraybackslash}p{1.3cm}>{\centering\arraybackslash}p{1.3cm}>{\centering\arraybackslash}p{1.3cm}>{\centering\arraybackslash}p{1.8cm}>{\centering\arraybackslash}p{1.8cm}}
	\hline\hline
	Impurity  & $\Delta\mu$ (eV) & $n(t_{2g})$ & $n(e_g)$ & $n(\mathrm{3d})$ & $Z(t_{2g})$ & $Z(e_g)$  & $T_\mathrm{K}(t_{2g})$ (K) & $T_\mathrm{K}(e_{g})$ (K)  \\
	\hline
    Ti   &  4.0 & 0.73 & 0.13 & 2.45 & 0.77 & 1.00 & 4171 & 5870  \\
    V &  3.7 & 1.12 & 0.15 & 3.66 & 0.67 & 0.79 & 3183 & 4003 \\
    Cr & 2.7 & 1.44 & 0.27 & 4.87 & 0.62 & 0.37 & 2649 & 1627  \\
    Mn & 1.1 & 1.40 & 0.76 & 5.72 & 0.35 & 0.22 & 1411 & 864 \\
    Fe & $-0.1$ & 1.60 & 1.06 & 6.93 & 0.40 & 0.087 & 1547 & 315  \\
    Co & $-1.9$ & 1.56 & 1.62 & 7.93 & 0.47 & 0.78 & 1833  & 2719 \\
    Ni & $-4.4$ & 1.80 & 1.82 & 9.03 & 0.98 & 0.97 & 3963 & 3314  \\
    \hline\hline
     \end{tabular}
\end{table}

The one-particle Hamiltonian in Eq.~\ref{eq:MAIM} was treated in a similar fashion as in the all-orbital calculations, where we adopted the Hartree-Fock effective Hamiltonian for five $3d$ orbitals so that the double-counting term can be exactly removed. The $3d$ block of the DFT hybridization function was discretized on the same non-uniform grid to obtain the bath parameters $\epsilon_k$ and $V_{ik}$, which resulted in 49 bath orbitals per $3d$ impurity orbital. 

To solve the embedding problem consisting of 5 impurity orbitals and 245 bath orbitals, we employed the same active-space DMRG solver. A Hartree-Fock calculation with fixed chemical potential at the DFT level was first performed, followed by a CISD calculation on the full embedding problem. A (40e, 40o) natural-orbital active space was derived by diagonalizing the CISD density matrix. We then conducted ground-state DMRG calculations with bond dimension $M=3500$ on the (40e, 40o) active space and further derived a (20e, 20o) DMRG natural-orbital active space. Finally, a dynamical DMRG calculation was done on the (20e, 20o) active space with bond dimension $M=1200$, and DMRG self-energies were used to estimate the quasiparticle renormalizations and Kondo temperatures in magnetic impurities. 

We summarize \REV{the Green's function results obtained using the same chemical potential as in the all-orbital calculations} in Table~\ref{tab:tkmodel} and the ground-state properties in Table~\ref{tab:densitymodel}. We find that the natural occupancies in the downfolded model calculations are less \REV{close to singly occupied than} in the all-orbital calculations. In the meantime, the quasiparticle renormalization weights and the predicted Kondo temperatures are much higher in the model calculations compared to the all-orbital simulations. \REV{We also observe that the total $3d$ occupancies in the model differ from those in the all-orbital calculation by up to 1.27 electrons. Because the $3d$ occupancy can significantly impact the correlation strength and the Kondo temperature~\cite{jacob2015towards}, for a fair comparison with the all-orbital calculations, we also performed the model calculation with chemical potentials $\mu$ fitted such that the total $3d$ occupancies match those in the all-orbital calculations in Table~\ref{tab:density}. The model results with the fitted $\mu$ are shown in Table~\ref{tab:fittedmodel}. The Kondo temperatures calculated using the new quasiparticle renormalization weight decrease slightly for a few elements (e.g., Fe and Co) compared to without the chemical potential fitting, but there is no consistent improvement in the absolute $T_\mathrm{K}$ or its trends.
With or without fitting the chemical potential in the low-energy model calculations, our all-orbital results are an order of magnitude better.}

\section{Specific heat calculation on the Anderson impurity model}

\label{sec:SIAMCv}
\REV{We benchmark the accuracy of a specific heat calculation on the symmetric single-impurity Anderson model (SIAM) (see \ref{sec:SIAM} for an introduction) within the numerical framework of this work. In the symmetric SIAM, for large $\frac{U}{\Delta_0}$, $T_\mathrm{K}$ can be calculated analytically as a function of the impurity on-site Coulomb interaction $U$ and the hybridization strength $\Delta_0$, $T_\mathrm{K}=\sqrt{\frac{U\Delta_0}{2}} \exp \left( - \frac{\pi U}{8\Delta_0} + \frac{\pi \Delta_0}{2 U}\right)$ ~\cite{Hewson1993a}. Additionally, the specific heat has been solved exactly through Bethe Ansatz and computed numerically~\cite{wiegmann1983exact, okiji1983thermodynamic}. This allows us to analyze the errors due to the bath discretization and active space truncation on the peak temperature of the specific heat and also verify the nature of the excitations that give rise to the Kondo singlet breaking (see \ref{sec:ftCv}). }

\REV{To compute the finite temperature properties, we solve for the excited states of the SIAM in the canonical ensemble using the ground-state DMRG solver with SU(2) symmetry and a level-shifted Hamiltonian for the $n^\mathrm{th}$ eigenstate
\begin{equation}
    \hat{H}^{\prime}_n = \hat{H} + \sum_{i=1}^{n-1} c_i |\Psi_i\rangle \langle\Psi_i|
\end{equation}
where $\hat{H}$ is the original DMRG Hamiltonian, $c_i$ is the energy shift multiplier, and $|\Psi_i\rangle$ is the $i^\mathrm{th}$ converged eigenstate to be projected out~\cite{Zhai2023a, wouters2014chemps2, fishman2022itensor}. We used $c_i=5D$ for all $i$ and a DMRG bond dimension of $M= 1200$. With this bond dimension, DMRG converged to a small discarded weight of less than $5\times10^{-9}$.}

\begin{figure*}[hbt!]
\centering
\includegraphics[width=0.5\textwidth]{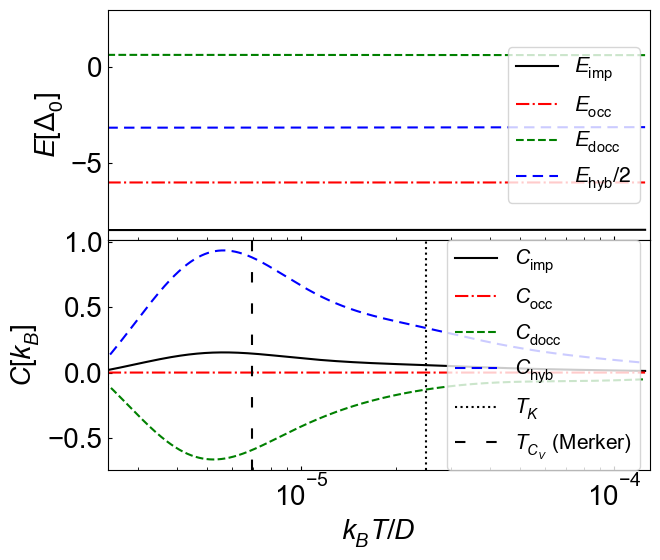}
\caption{\REV{Benchmark of finite temperature results from the DMRG solver on the SIAM with $U=0.012 D$ and $\Delta_0 = 0.001 D$, using the logarithmic discretization with base $b=4$ and the procedure to extract impurity specific heat as outlined in Ref.~\cite{merker2012numerical}. Different components of the energy and specific heat are defined the same as in Fig.~4 of Ref.~\cite{merker2012numerical}. The exact $T_\mathrm{K}$ (vertical black dotted line) and specific heat peak temperature from numerical renormalization group results in Ref.~\cite{merker2012numerical} (vertical black dashed line, ``Merker'') are included as references.}}
\label{fig:merker}
\end{figure*}

\REV{We benchmarked on the SIAM with the same parameters as in Ref.~\cite{merker2012numerical}, $U=0.012 D$ and a flat-band hybridization $\mathrm{Im} \Delta(\omega+i0^+)=-\Delta_0, ~ |\omega| < D$ where $\Delta_0 = 0.001 D$. For convenience, we set $D=1$. We used the same logarithmic discretization, consistent with the rest of the manuscript, but with 19 orbitals and logarithmic intervals defined using a larger log base $b=4$  
as used in Ref.~\cite{merker2012numerical} to achieve a small spacing near the Fermi level. Using the procedure to extract the impurity specific heat contribution as outlined in Ref.~\cite{merker2012numerical}, }
\REV{we obtained a peak in the impurity specific heat at 0.22 $T_\mathrm{K}$ (Fig.~\ref{fig:merker}), in good agreement with the peak in Ref.~\cite{merker2012numerical} at around 0.28 $T_\mathrm{K}$.}

\begin{figure*}[hbt!]
\centering
\includegraphics[width=0.5\textwidth]{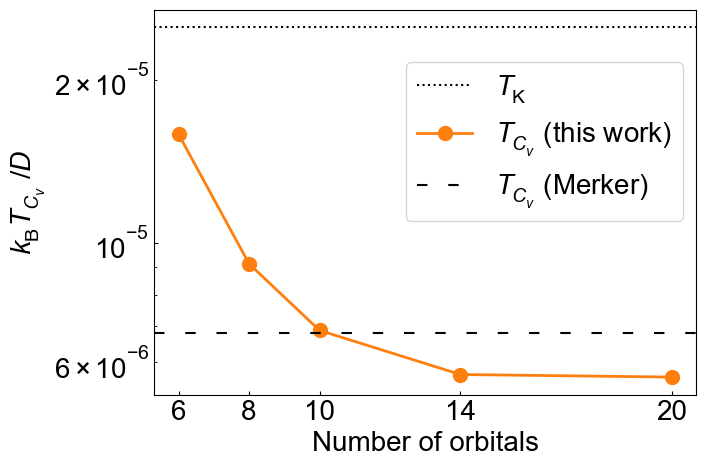}
\caption{\REV{Specific heat peak temperature ($T_{C_\mathrm{V}}$)  of SIAM with $U=0.012 D$ and $\Delta_0 = 0.001 D$, using active spaces of different numbers of orbitals taken out of the full space of 20 orbitals, the logarithmic discretization with base $b=4$ and the procedure to extract the impurity specific heat as outlined in Ref.~\cite{merker2012numerical}. The exact $T_\mathrm{K}$ (black dotted line) and specific heat peak temperature from numerical renormalization group results in Ref.~\cite{merker2012numerical} (black dashed line) are shown as references.}}
\label{fig:merkeras}
\end{figure*}

\REV{In Fig.~\ref{fig:merkeras} we show the estimated specific heat peak temperatures $T_{C_\mathrm{V}}$  as a function of the (DMRG natural orbital) active space size. We see that the position of the peak varies by about a factor of 3 as a function of the active space size (and thus the estimated Kondo temperature also varies by about a factor of 3). This is different from the behaviour of the estimated Kondo temperatures from the Hewson formula, as shown in \ref{sec:SIAM}, which are already reasonably converged with an active space size of 8 orbitals. Since the location of the $C_\mathrm{V}$ peak depends on different excitations (in particular the neutral excitations) than those that enter into the self-energy and $Z$, it is unsurprising that there are different finite size effects. These numerical results support the use of $T_\mathrm{K}$ from the Hewson formula as our primary estimate in the main text.}

\REV{From the excited states, we can also examine the nature of the excitations that give rise to the peak in the heat capacity. To verify the spin-flip nature of the lowest energy excitations, we compute the
impurity spin-flip weight $w_{ii}^2 = |\langle \Psi_{S=0} | c^\dag_{i\beta} c_{i\alpha}   | \Psi_{S=1}\rangle|^2$ where $i$ indexes the impurity orbital (here, the $S_z=1$ state of the $|\Psi_{S=1}\rangle$ multiplet is used as an example). This quantity satisfies $w_{ii}^2 \leq 1$ and is equal to 1/2 for a pure localized impurity spin flip. We show the weights for the lowest excited states, with their corresponding energies, in Table~\ref{tab:siamexcitations}.}

\begin{table}[hbt!]
	\centering
	\caption{\REV{Impurity spin-flip weight $w^2_{ii}$, its normalized value $w^2_{ii}/\sum_{pq} w^2_{pq}$, and the excitation energies $E/D$ of the lowest singlet-triplet excitations that give rise to the specific heat peak, calculated in active spaces of different number of orbitals ($N_\mathrm{orb}$).}}
	\label{tab:siamexcitations}
	\begin{tabular}{>{\centering\arraybackslash}p{2cm}>{\centering\arraybackslash}p{3cm}>{\centering\arraybackslash}p{3cm}>{\centering\arraybackslash}p{2cm}>{\centering\arraybackslash}p{2cm}>{\centering\arraybackslash}p{2cm}}
	\hline\hline
	$N_\mathrm{orb}$   & threshold (occupied) & threshold (virtual) & $w^2_{ii}$  & $w^2_{ii}/\sum_{pq} w^2_{pq}$ & $E/D$  \\
	\hline
    20 & 0 & 0 & 0.071 & 0.123 & $1.2\times 10^{-5}$\\
    14 & $1\times10^{-7}$ & $1\times10^{-7}$ & 0.073 & 0.105 & $1.2\times 10^{-5}$\\
    10 & $3\times 10^{-5}$ & $3\times 10^{-5}$ & 0.087 & 0.138 & $1.5\times 10^{-5}$\\
    8  & $2\times 10^{-4}$ & $2\times 10^{-4}$ & 0.101 & 0.169 & $2.0\times 10^{-5}$\\
    6  & $2\times 10^{-3}$ & $2\times 10^{-3}$ & 0.121 & 0.210 & $3.4\times 10^{-5}$\\
    \hline\hline
     \end{tabular}
\end{table}

\section{Ab initio impurity specific heat}\label{sec:ftCv}

\REV{We calculated the finite-temperature and excited state properties in the all-electron simulations within different active spaces, using the procedure outlined in \ref{sec:SIAMCv}.
We used a grand canonical formalism and extracted the canonical $C_\mathrm{V}$ using the relations in Ref.~\cite{stroker2021classical}.
The chemical potential was chosen to ensure that the average electron number of the impurity and bath at $T_\mathrm{K}$ (estimated from Hewson's formula Eq.~\ref{eq:hewson}) matched the electron number of the ground state. The excited states were obtained in a DMRG active space of up to (26e, 36o) size with  maximum DMRG bond dimension $M= 4000$, converging to a discarded weight of less than $5\times10^{-5}$. 
In this way, 449, 129, 83, 44, 42, 38, and 80 excited states in total were calculated for the Ti, V, Cr, Mn, Fe, Co, and Ni impurities, respectively.}

\REV{In the SIAM, the impurity contribution to the specific heat was obtained by partitioning the energy between the impurity and the bath, as described in Ref.~\cite{merker2012numerical}. However, in the case of the ab initio model, the partitioning is ambiguous. This is in part because energy partitioning is always non-unique (as the only constraint is that the parts sum up to the same whole), but also because as we converge to the basis limit of the all-electron basis, the additional impurity orbitals necessarily extend beyond the impurity atom into the bulk (see discussion in main text). We found that this either made the estimated $C_\mathrm{V}$ sensitive to various choices, or gave unphysical results (e.g., the impurity specific heat capacity could be negative).}

\begin{figure*}[hbt!]
\centering
\includegraphics[width=\textwidth]{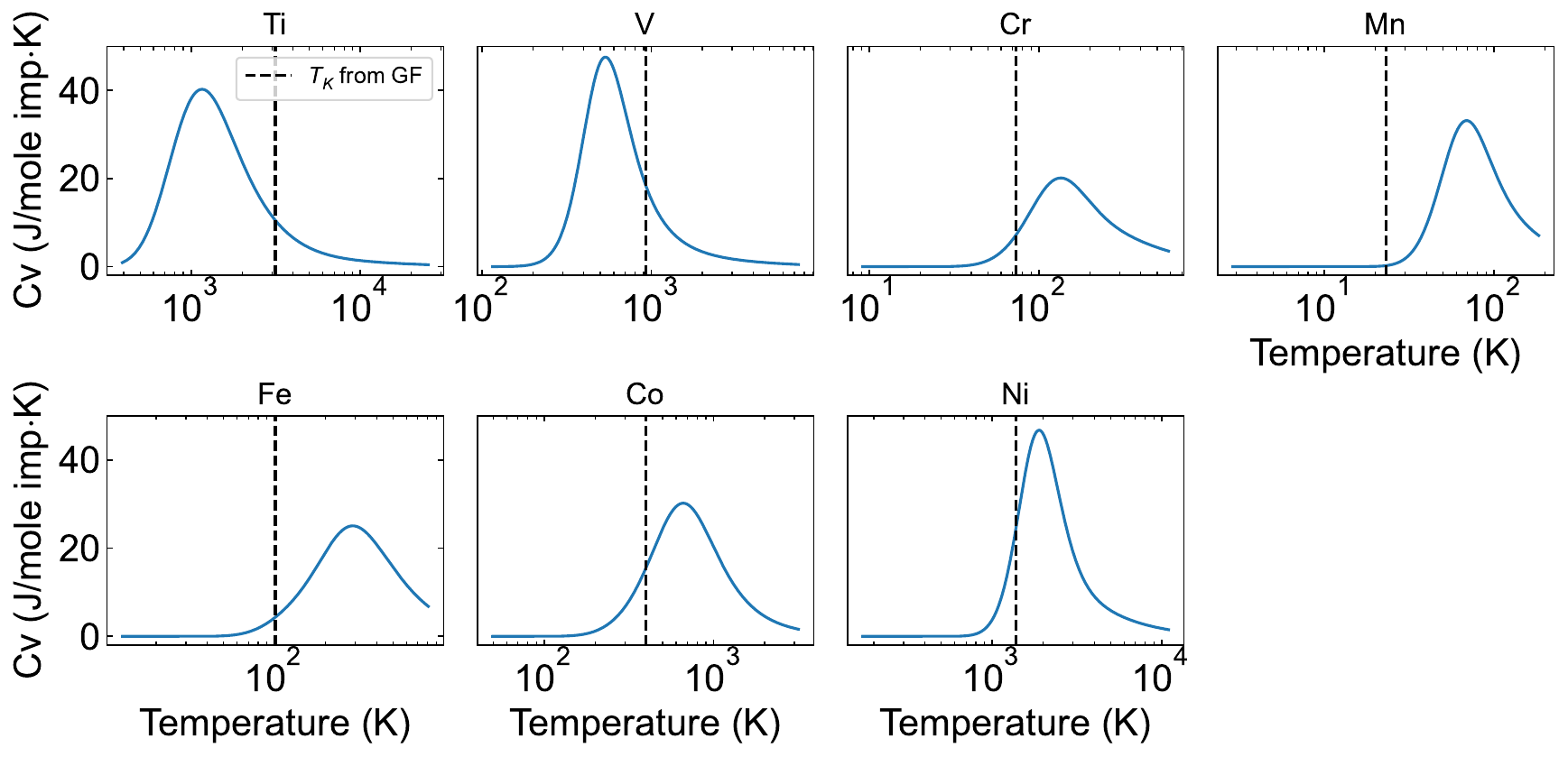}
\caption{\REV{Impurity specific heat per mole of magnetic impurities in bulk Cu from ab initio all-orbital calculations. The Kondo temperatures predicted using Herson's formula with the quasiparticle renormalization weights from the Green's function calculations (``$T_\mathrm{K}$ from GF'') are shown for comparison.}}
\label{fig:Cv}
\end{figure*}

\begin{figure*}[hbt!]
\centering
\includegraphics[width=0.6\textwidth]{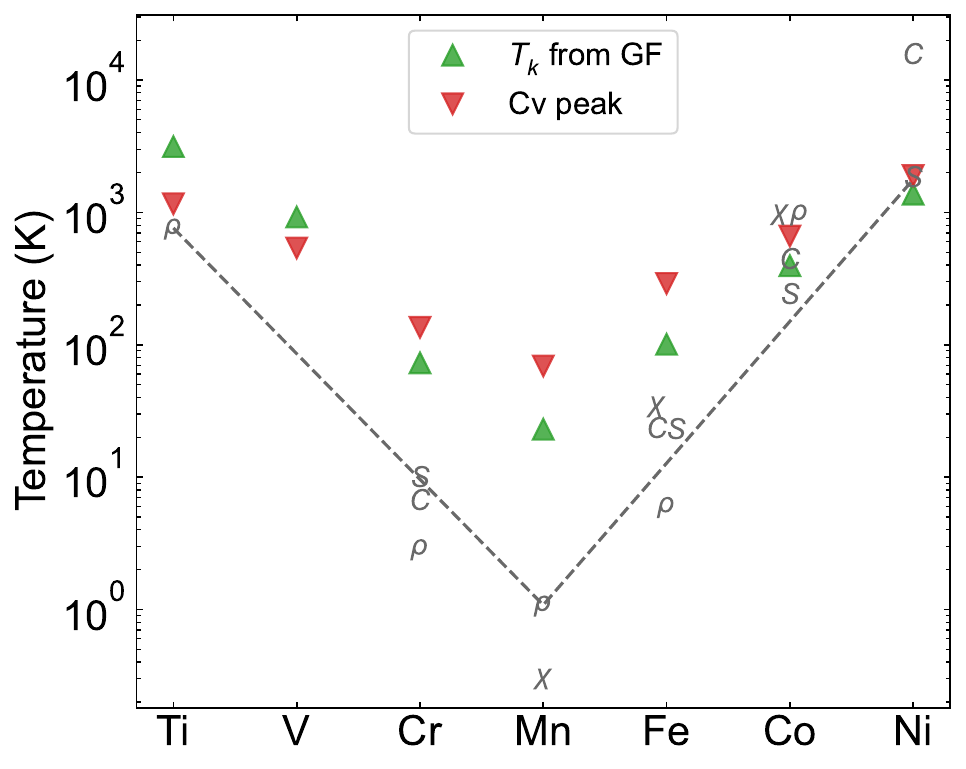}
\caption{\REV{The temperatures corresponding to the peak in the impurity specific heat compared to the Kondo temperatures predicted by Hewson's formula from ab initio all-orbital calculations and those measured in experiments. }}
\label{fig:Cv_trend}
\end{figure*}

\REV{Consequently, to produce a (rough) estimate of $T_\mathrm{K}$, similar to the analysis of the excitation character of the excited states in \ref{sec:SIAMCv}, we quantified the degree of ``impurity'' character of each excited state and only included those above a given threshold when computing the impurity specific heat. Concretely, for the $S_z=1$ component of a triplet excited state, we
computed
\begin{equation}
    w_{ij} = \langle \Psi_0 | c^\dag_{i\alpha} c_{j\beta} | \Psi_1\rangle
\end{equation}
where $c^\dag_{i\alpha}, c_{j\beta}$ are creation and annihilation operators on orbitals that are localized on the impurity or on the bath, respectively (these orbitals were obtained by a rotation of the active space). Then, the impurity character is defined as
\begin{equation}
    \lambda = \frac{\sum_{i \notin \mathrm{bath} \ \mathrm{or} j \notin \mathrm{bath}} w_{ij}^2} { \sum_{ij} w_{ij}^2}.
\end{equation}
A similar analysis was carried out for the quintet excited states. Using a threshold of $\lambda > 0.12$, we show the estimated $T_{C_\mathrm{V}}$ from the impurity specific heat peak in Fig.~\ref{fig:Cv}. In the SIAM, the peak of the specific heat is always below $T_\mathrm{K}$. We see that this is true in our calculations for Ti and V, but not for the other impurity elements, suggesting that $T_\mathrm{C_\mathrm{V}}$ is overestimated in the other cases. Given the strong finite size effects observed in specific heat capacity calculations in general (as studied in the SIAM in \ref{sec:SIAMCv}), we consider the estimation of $T_\mathrm{K}$ from the Hewson formula to be better converged. However, this rough treatment indicates that similar trends can be reproduced through the heat capacity (see Fig.~\ref{fig:Cv_trend}).}

\section{Ab initio and model excited state analysis}

\begin{figure*}[hbt!]
\centering
\includegraphics[width=\textwidth]{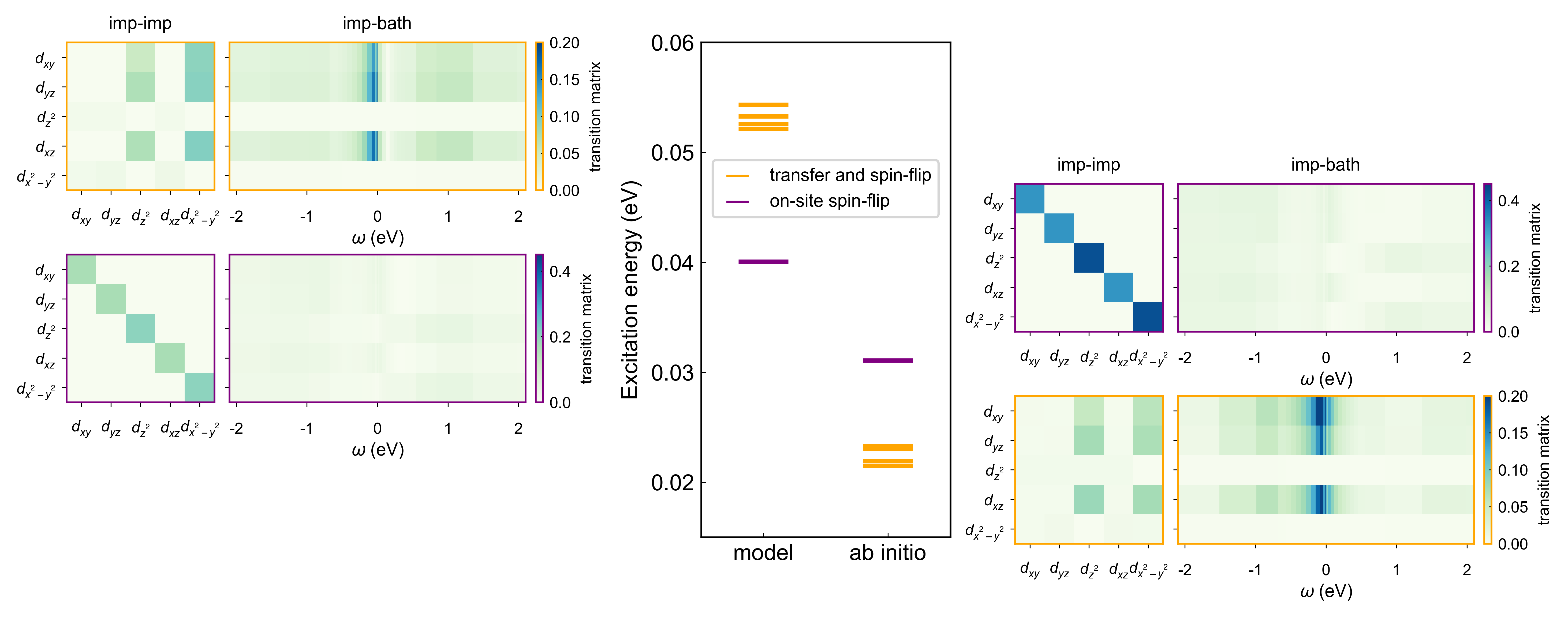}
\caption{\REV{The excitation energies of triplets associated with on-site spin-flip (purple) from the ground state and of triplets associated with charge transfer between different orbitals and spin-flip (orange) from the ground state for the Mn impurity. The impurity-impurity and impurity-bath blocks of spin-flip transition density matrices $|\langle \Psi_{S=0} | c^\dagger_{i\downarrow} c_{j\uparrow} | \Psi_{S=1, S_z=1}\rangle|$ from the ground state (vertical axis) to the triplet(s) (horizontal axis) from model calculations and \textit{ab initio} all-orbital calculations are shown on the left and right, respectively, to illustrate the on-site spin-flip (purple) and the charge transfer and spin-flip (orange) process. The bath orbitals are ordered based on their bath energies $\omega$. The transition density matrices of charge-transfer triplets are summed over all near-degenerate states.  }}
\label{fig:transfer}
\end{figure*}

\REV{To illustrate the difference in the excited states in the ab initio all electron treatment and in the model calculations, we performed excited state model calculations for Mn in the (20e, 20o) active space with the same chemical potential as in the all-orbital calculations, and compared those to the excited states seen in the ab initio calculations. As shown in Fig.~\ref{fig:transfer}, in both the model and \textit{ab initio} all-orbital calculations, we observed multiple near-degenerate low-energy triplets that involve partial transfer of an electron from $t_\mathrm{2g}$ to $e_g$ and to the bath orbitals, accompanied by a spin-flip (orange), as well as one low-energy triplet that corresponds to a simultaneous on-site spin-flip of all fractionally occupied $d$ orbitals (purple). The latter corresponds to the low-energy excitation of a large spin Kondo model. However, in the model calculations, the charge transfer triplets are higher in energy than the Kondo model triplet, which is opposite to the all-orbital calculations where the charge-transfer triplets are the lowest triplets. As discussed in the main text, correctly capturing the charge-transfer triplet excitations appears to be critical to reproducing the quantitative Kondo trends, which highlights again the power of the all-orbital calculations.}

\begin{figure*}[hbt!]
\centering
\includegraphics[width=0.8\textwidth]{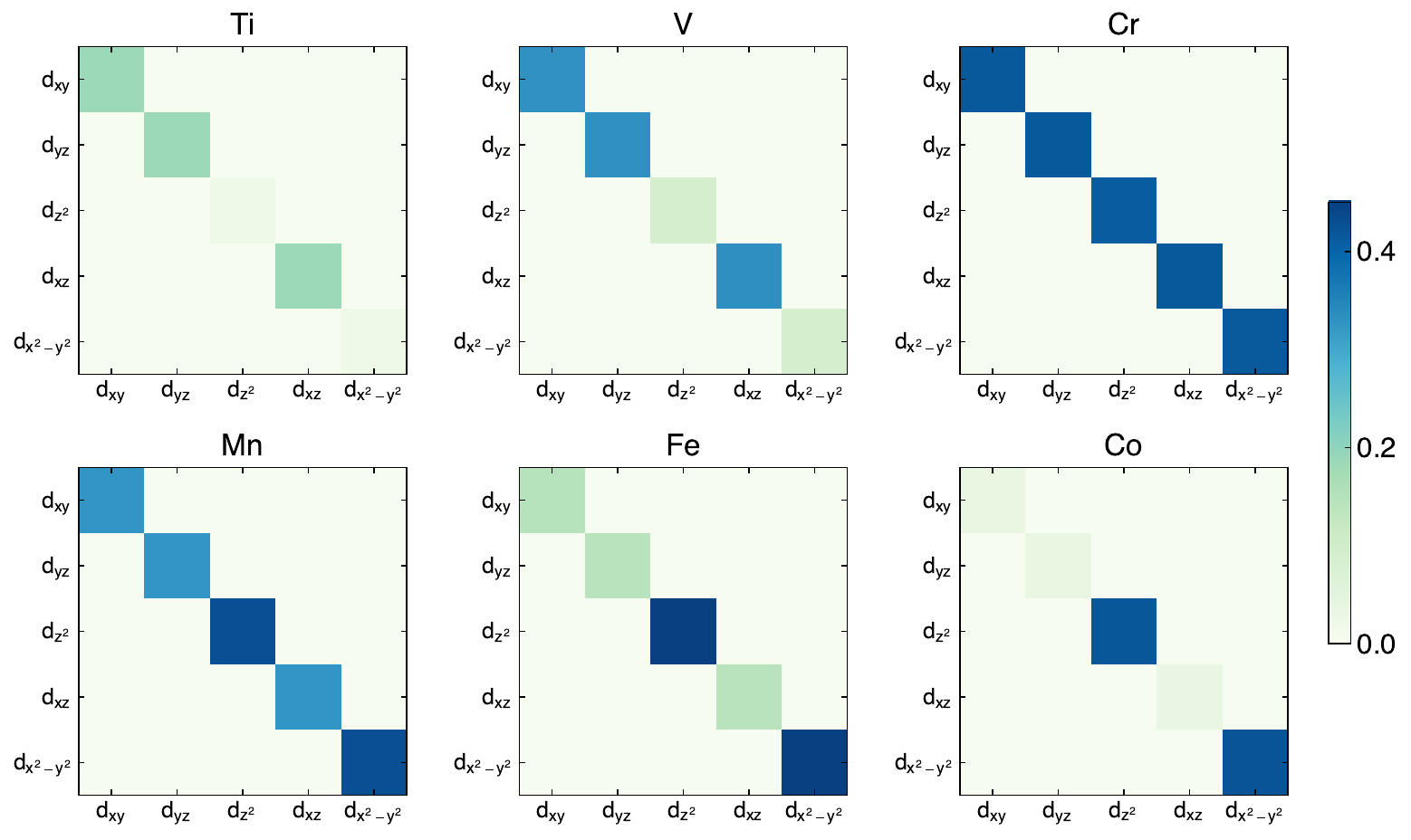}
\caption{\REV{Orbital resolved spin-flip transition density matrix  $|\langle \Psi_{S=0} | c^\dagger_{i\downarrow} c_{j\uparrow} | \Psi_{S=1, S_z=1}\rangle|$ between the ground state and the lowest triplet excited state that is dominated by intra-impurity orbital excitations, across the series of impurities from Ti to Co.}}
\label{fig:tdm}
\end{figure*}

\REV{Fig.~\ref{fig:tdm} further shows the ab initio transition density matrix between the ground-state and the low-energy triplet state which is dominated by a pure intra-impurity spin-flip for all the elements, corresponding to the excitation in the multi-orbital Kondo model. For Ti and V, the singlet-triplet excitation is dominated by the $t_{2g}$ intraorbital excitations, while for Fe and Co it is dominated by the $e_g$ intraorbital excitations. The excitations in Cr and Mn have significant contributions from both $t_{2g}$ and $e_g$ orbitals.}

\FloatBarrier

\section{Computational time}

\begin{table}[h!]
	\centering
	\caption{\REV{Computational time of various tasks: ground state DMRG calculations in the largest active space (36e, 52o); Green's function calculations with DDMRG at 7 frequency points near the Fermi level in active spaces of around (22e, 22o) ((22e, 27o) for Co); and specific heat ($C_V$) calculations where we calculated ground and excited eigenstates in spin manifolds of up to $S=8$ in the canonical ensemble in active spaces of around (28e, 28o) ((30e, 29o) for Ti, (28e, 30o) for V, (26e, 36o) for Co).  The ground state and Green's function calculations were performed on 2--8 nodes with 24 cores per node, while the specific heat calculations in each spin manifold were performed on 1--8 nodes with 64 cores per node.   }}
    
	\label{tab:timing}
	\begin{tabular}{>{\centering\arraybackslash}p{1.7cm}>{\centering\arraybackslash}p{1.9cm}>{\centering\arraybackslash}p{1.9cm}>{\centering\arraybackslash}p{1.9cm}>{\centering\arraybackslash}p{1.9cm}>{\centering\arraybackslash}p{1.9cm}>{\centering\arraybackslash}p{1.9cm}}
	\hline\hline
    & \multicolumn{2}{c}{\textbf{Ground state (36e, 52o)}} & \multicolumn{2}{c}{\textbf{Green's function (22e, 22o)}} 
    & \multicolumn{2}{c}{\boldsymbol{$C_V$}  \textbf{(28e, 28o)}} \\
	\cline{2-3} \cline{4-5} \cline{6-7}
    Impurity  & wall time (h) & core hours & wall time (h) & core hours  & wall time (h) & core hours \\
	\hline
    Ti   &  27 & 2556 & 60 & 5776 & 1019 & 119784 \\
    V    &  55 & 7984 & 94 & 18057 & 733 & 95274 \\
    Cr   &  64 & 9215 & 99 & 18959 & 645 & 50843 \\
    Mn   &  100 & 14395 & 105 & 20088 & 489 & 49726 \\
    Fe   &  82 & 11853 & 135 & 25848 & 237 & 15174 \\
    Co   &  78 & 11225 & 298 & 71534 & 171 & 21884 \\
    Ni   &  49 & 9407 & 66 & 12753 & 50 & 3179 \\
    \hline\hline
     \end{tabular}
\end{table}
\pagebreak

\section{Kondo temperature in multi-orbital impurities}
\REV{The exponential decrease in the Kondo temperature with the spin $S$ of the impurity~\cite{Daybell1968} has previously been discussed using a mechanism involving  Hund's coupling~\cite{Schrieffer1967}. We briefly summarize the literature arguments here.}

\REV{As a simple example, consider an Anderson model with  $l=2$  conduction electrons scattering off an impurity with a half-filled $d$-shell,
\begin{equation}
H=\sum_{\boldsymbol{k}m\sigma}\epsilon_{\boldsymbol{k}} n_{\boldsymbol{k}m\sigma} + \sum_{\boldsymbol{k} m \sigma} V_{\boldsymbol{k} m} \left(c^{\dagger}_{\boldsymbol{k}\sigma}f_{m\sigma}+ f^{\dagger}_{m\sigma}c_{\boldsymbol{k}\sigma}\right) + H_\mathrm{imp} \text{,}
\end{equation}
where $H_\mathrm{imp}$ includes the interaction on impurity sites and $m, \sigma$ are the labels of the atomic orbital and spin separately.}

\REV{In the presence of strong on-site electron repulsion and Hund's coupling, we can focus solely on the spin degree of freedom of the impurity. The spin configuration of the impurity is constrained to  $S = 5/2$ . Using the Schrieffer-Wolff transformation~\cite{Schrieffer1966, Schrieffer1967}, the effective Hamiltonian is reduced to the following form:
\begin{equation}
H = \sum_{\boldsymbol{k}m\sigma}\epsilon_{\boldsymbol{k}} n_{\boldsymbol{k}m\sigma} + \sum_{\boldsymbol{k}\boldsymbol{k}^{\prime}m} J_{m\boldsymbol{k}\boldsymbol{k}^{\prime}} 
\boldsymbol{S} \cdot c^{\dagger}_{\boldsymbol{k}^{\prime}m} \frac{\boldsymbol{\sigma}}{2} c_{\boldsymbol{k}m} \text{.}
\end{equation}
The orbital angular momentum of the scattering conduction electrons is conserved. Because the impurity is half-filled, conduction electrons cannot exchange orbital angular momentum with the impurity. The exchange interaction  $J_{m\boldsymbol{k}\boldsymbol{k}^{\prime}}$ is second order in the hybridization term:
\begin{equation}
J_{m\boldsymbol{k}\boldsymbol{k}^{\prime}}= \frac{V_{\boldsymbol{k}m} V_{\boldsymbol{k}^{\prime}m}}{2S}(\frac{1}{E_+} + \frac{1}{E_{-}})\text{,}
\label{J-dep}
\end{equation}
where $E_{+}$ and $E_{-}$ represent the additional energy of the virtual process involving the addition or removal of an electron. The factor of $2S$ in the denominator arises from the representation of the $S=5/2$ spin operator in terms of five $S=1/2$ impurity electron operators.}

\REV{For an impurity without a half-filled $d$-shell, the effective Hamiltonian includes both orbital and spin exchange. However, if we focus only on the spin degree of freedom and assume that all singly occupied states are spin-up, with the remaining states either doubly occupied or empty, the spin dependence of $J_{m\boldsymbol{k}\boldsymbol{k}^{\prime}}$ can be generalized, where spin corresponds to the spin of the singly occupied orbitals~\cite{Schrieffer1967}. Although simplistic, this formula provides a rationalization for the trend of the Kondo temperature $T_\mathrm{K}$ of transition metal impurities within host alloys~\cite{Daybell1968,Nevidomskyy2009} with 
\begin{equation}
T_\mathrm{K} \sim \exp(-1/J^\mathrm{eff} \rho) \text{.}
\end{equation}
Here, $J^\mathrm{eff}$ is inversely proportional to the spin $S$ as in Eq.~\ref{J-dep}, and consequently, the Kondo temperature is predicted to exhibit an exponential dependence on $S$.}

\REV{Another way to estimate the Kondo temperature is through the bandwidth of the Kondo resonance (Abrikosov-Suhl resonance),  $\tilde{\Delta}$ , at the Fermi level, with
\begin{equation}
T_k = \frac{\pi}{4} \tilde{\Delta} \text{.}
\end{equation}
If there is no quantum phase transition between the non-interacting limit and the finite-interaction case, the low-energy behavior of the interacting system can be described by a local Fermi liquid theory~\cite{Nozieres1974}. The bandwidth $\tilde{\Delta}$ in the interacting case is related to the non-interacting hybridization width  $\Delta$  through the quasiparticle renormalization weight $Z$ :
\begin{equation}
\tilde{\Delta} = Z \Delta \text{.}
\end{equation}
The entire effect of the interaction is encapsulated in $Z$. If the renormalization weight $Z$ can be estimated non-perturbatively, this formula can be applied to address the finite interaction case, which is the basis of the Hewson estimate of the Kondo temperature used in the main text.}

\printbibliography[title={References in Supplementary Materials}, notkeyword=appendix]
\end{refsection}

\end{document}